\documentclass[10pt,journal,twocolumn]{IEEEtran}

\usepackage[cmex10]{amsmath}
\usepackage{amsmath, amsthm, amssymb, cite, graphicx, algorithm, epsfig, url, color, array, subfigure}
\usepackage[noend]{algpseudocode}
\numberwithin{algorithm}{section}
\newcommand{\ignore}[1]{}

\newtheorem{proposition}{Proposition}[section]

\newtheorem{lemma}{Lemma}[section]

\usepackage[belowskip=-5pt,aboveskip=5pt]{caption}
\begin{document}


\title{Ensuring Predictable Contact Opportunity for Scalable Vehicular Internet Access On the Go}

\author{Zizhan~Zheng,~\IEEEmembership{Member,~IEEE,} Zhixue~Lu,~\IEEEmembership{Student Member,~IEEE,} Prasun~Sinha,~\IEEEmembership{Senior Member,~IEEE,} and~Santosh~Kumar,~\IEEEmembership{Senior Member,~IEEE}%
\thanks{Z. Zheng is with Department of Electrical and Computer Engineering,
The Ohio State University. E-mail: zhengz@ece.osu.edu.}%
\thanks{Z. Lu is with Department of Computer Science and Engineering,
The Ohio State University. E-mail: luz@cse.ohio-state.edu.}%
\thanks{P. Sinha is with Department of Computer Science and Engineering,
The Ohio State University. E-mail: prasun@cse.ohio-state.edu.}%
\thanks{S. Kumar is with Department of Computer
Science, University of Memphis. E-mail: santosh.kumar@memphis.edu}%
\thanks{A preliminary version of this work titled ``Maximizing the Contact Opportunity for Vehicular Internet Access" appeared in the Proceeding of the IEEE INFOCOM 2010~\cite{contactopp}.}}

\maketitle

\begin{abstract}
With increasing popularity of media enabled hand-helds and their integration with the in-vehicle entertainment systems, the need for high data-rate services for mobile users on the go is evident. This ever-increasing demand of data
is constantly surpassing what cellular networks can economically support. Large-scale
Wireless LANs (WLANs) can provide such a service, but they are
expensive to deploy and maintain. Open WLAN access-points, on
the other hand, need no new deployments, but can offer only
opportunistic services, lacking any performance guarantees.
In contrast, a carefully planned {\it sparse} deployment of roadside
WiFi provides an economically scalable infrastructure with quality
of service assurance to mobile users. In this paper, we present a new metric, called {\it Contact
Opportunity}, to closely model the quality of data service that a mobile user might experience
when driving through the system. We then present
efficient deployment algorithms for minimizing the cost for ensuring a required level of contact
opportunity. We further extend
this concept and the deployment techniques to a more intuitive
metric -- the average throughput -- by taking various dynamic
elements into account. Simulations over a real road network and experimental results show that our approach achieves significantly better cost vs. throughput tradeoff in both the worst case and average case
compared with some commonly used deployment algorithms.
\end{abstract}

\section{Introduction}\label{sec:intro}

With increasing popularity of media enabled hand-helds and their integration with the in-vehicle entertainment systems, the need for
high data-rate services for mobile users on the go is growing rapidly. This ever-increasing demand of data
is routinely surpassing what cellular networks can economically support.
WiFi hotspots are rapidly mushrooming in every city to meet this demand. They either operate independently as a competitive way of data access, or act as a complementary service and help offload the overburdened 3G networks~\cite{wifirescue}. But, their primary target is static users. These networks fail to provide
any assured level of service to a mobile user. Although large
deployments of WLANs can be used to provide high data-rate services
over large areas, the cost becomes prohibitive due to the sheer
number of access-points (APs) required. For instance, to cover a 2km
x 2km area in Mountain View, Google needed to deploy 400 access
points \cite{googlewlan} to barely provide coverage at the base data
rate. In addition to the deployment cost, the maintenance and
management complexity has led to abandonment or scaling back of
several WLAN projects from San Francisco to Philadelphia
\cite{BWEEK}.

\begin{figure}
\centering
\includegraphics[width=2.5in]{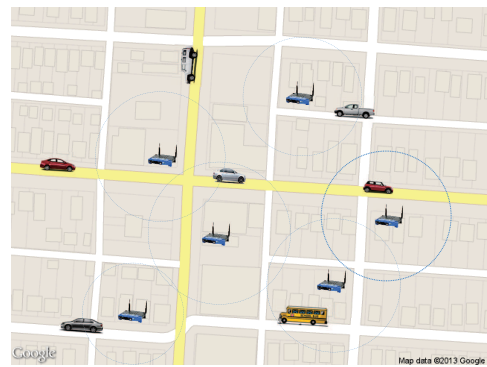}
\caption{\small Vehicular Internet access via roadside WiFi } \label{fig:road}
\end{figure}

New Wireless Wide-Area Networking (WWAN) technologies such as 3GPP
LTE (Long Term Evolution) and mobile WiMAX are expected to provide
{\it either} long range coverage {\it or} high data rates, but
practical numbers are far from the promised levels. For example
WiMAX is intended to support data rates as high as 75 Mbps per 20
MHz channel, or a range of 30 miles \cite{wimaxrange}. However, one
of the first deployments of WiMAX in US is reported to provide a
downlink bandwidth of 3 Mbps \cite{BWEEK}, which is only within a
factor of 2 better than the current 3G networks. Note that these
resources will potentially be shared by a large number of active
users within the respective sector of the antenna. Given the
resistance from majority of users to pay high monthly fees for
mobile data access, which is essential for supporting expensive new
deployments, ubiquitous service from such new deployments could take
several years, and possibly decades.

On the other hand, evaluation of wireless data access by mobile
users using ``in situ" (or ``open") WiFi
networks~\cite{CarTel,MobiSteer,Interactive,Cabernet}, and in
various controlled
environments~\cite{drive-thru,in-motion,MobiSteer,Interactive} have
confirmed the feasibility of WiFi based vehicular Internet access
for non-interactive applications. The possibility and challenges to
support certain interactive applications, such as Web browsing, have
also been studied~\cite{websearch,Interactive}. Most existing works,
however, consider an unplanned deployment of APs based on open-APs
\cite{drive-thru,CarTel,MobiSteer,Interactive,Cabernet,in-motion}.
Consequently, these solutions {\it fail to provide any throughput
assurance} to a mobile user; they can only provide opportunistic
services to mobile users.

The two objectives -- an economically scalable infrastructure and
quality of service assurance -- can be achieved by a carefully
planned {\it sparse} deployment of WiFi APs at roadside. In this paper,
we study deployment techniques for providing roadside WiFi
services. We envision a wireless service provider that implements a deployment using two types of APs, new APs that are deployed for serving mobile users exclusively, and existing APs that are incentivized for sharing their capacity between static and mobile users. It is likely that these existing APs are initially deployed for serving static users or users with limited mobility and are possibly owned by other service providers or end users, and therefore will give higher priority to their original, mostly static, users.

To provide guaranteed performance to mobile users, we present a new metric, called {\bf
Contact Opportunity}, as a characterization of a roadside WiFi
network. Informally, the contact opportunity for a given deployment
measures the fraction of distance or time that a mobile user is in
contact with some APs when moving through a certain trajectory. Such a
metric is closely related to the quality of data service that a
mobile user might experience while driving through the system.
Our objective is to find a deployment that ensures a required level of contact opportunity with the minimum cost. Since the problem is NP-hard, we have designed an efficient approximation solution by exploiting a diminishing return property in the objective function.
We further show how to extend this concept and the deployment techniques to a more
intuitive metric -- the average throughput -- by taking various
dynamic elements into account. In particular, we take an interval based approach to model the uncertainties associated with road traffic conditions and the time varying data traffic load of static users.
The deployment algorithm is then extended to achieve a required level of average throughput under uncertainties, where we consider both a robust optimization approach that minimizes the cost in the worst-case scenario, and a two-stage stochastic optimization approach that minimizes the expected cost.

While focusing on WiFi deployment, our study also provides useful insights to the large deployment of other types of wireless networks, such as femotcells, for serving mobile users. Femotocells are small cellular base-stations initially designed to improve the indoor cellular coverage. But they are currently being extended to provide high data-rate coverage over short ranges to the outdoor environment as well~\cite{QualcommSmallCell,UbiquisysSmallCell}, and can potentially be utilized to support data-incentive services for mobile users. Our techniques can be applied to deploying new femotocell base-stations (FBSs) as well as acquiring service from existing FBSs that originally target at static users. One challenge to achieve a scalable infrastructure for serving mobile users using FBSs is to properly model the dynamics of data traffic load associated with both femtocells and macrocells.

This is the first work that addresses the challenges in achieving a {\it sparse}
wireless infrastructure that provides QoS assurance to mobile users {\it in the face of uncertainty}. The deployment issues with respect to roadside WiFi
networks have not received much attention in the past. Our previous work on Alpha
Coverage~\cite{alphacover,alphacover-journal} initiated research in this area.
In Alpha Coverage, each AP is viewed as
a point, and the objective is to bound the gap between two consecutive contacts with APs
when moving through a road network.
In other words, only the \emph{number} of contacts is
considered but not the \emph{quality} of each contact. In contrast,
contact opportunity is more closely related to the real performance
that a mobile user experiences by taking various static and dynamic
parameters into account, such as the coverage region of an AP at each
potential location, driving speed, and the free capacity of APs. Consequently,
finding an optimal deployment in terms of contact opportunity is
significantly more challenging. Following our studies in~\cite{alphacover,alphacover-journal,contactopp}, a few works have also considered planned deployment of APs and Road Side Units (RSUs) for serving mobile vehicles~\cite{Malandrino-Content-TMC,Delay-Constrained-Coverage}. However, these works either assume perfect knowledge of vehicle mobility in both space and time and do not provide a scalable solution with performance guarantee~\cite{Malandrino-Content-TMC}, or focus on a single (expected) scenario of road traffic and ignore the capacity constraints of APs~\cite{Delay-Constrained-Coverage}. We note that minimizing the cost of the expected scenario is different from our objective of minimizing the expected cost, and the latter is usually more difficult (see Section~\ref{sec:two-stage} for a detailed explanation).

We make the following contributions in this paper.
\begin{itemize}
\item We present a metric, called Contact Opportunity, as a characterization of roadside WiFi
deployment, which is closely related to the quality of data service
that a mobile user might experience when driving through the
network.
\item We design an efficient deployment method that ensures a required level of contact opportunity at a minimum cost by utilizing submodular optimization techniques.
\item We extend the concept of contact opportunity and the deployment techniques to average throughput by taking various dynamic
elements into account, and propose algorithms for minimizing the worst-case cost and the expected cost, respectively.
\item Simulations over a real road network and experimental results show that our approach
achieves a much better cost vs. throughput tradeoff in both the worst case and the average case
compared with some commonly used deployment approaches.
\end{itemize}

In addition to serving the mobile users on the go for media services,
several other applications can benefit from a roadside WiFi network
deployed using our techniques. Remote monitoring and tracking of
shipments is one such application. For example, Walmart currently
depends on a satellite based system \cite{VERIWISE} for tracking its
trailers, which is an expensive solution.
Similarly, businesses with mobile workforce can benefit from
media-rich communication over such a system.
%
%
Recently, the feasibility and usefulness of a system that provides
road condition updates has been studied \cite{GPSNOKIA}, which is
another use case of the system. Our techniques can be applied to
either plan a new deployment or to improve an existing one. For
instance, a new set of WiFi APs can be added to a set of open APs to
improve the quality of service.

The rest of the paper is organized as follows. A summary of related work is provided in Section~\ref{sec:related}.
The formal definition of contact opportunity, the
deployment problem, and our solution for a given network scenario are discussed in Section
\ref{sec:acquire}. The extensions of contact opportunity to average throughput and algorithms for coping with uncertainties are discussed in Section~\ref{sec:extension}.
Numerical results and experiments are presented in Sections \ref{sec:sim} and
\ref{sec:exp}, respectively.
We conclude the paper in Section~\ref{sec:concl}. 
\section{Related Work}\label{sec:related}
\ignore{As discussed in Section~\ref{sec:intro}, leveraging idle
Internet access bandwidth of roadside Access Points (APs) promises
to provide an inexpensive alternative to new WLAN and WWAN
deployments, that are expensive to both build and maintain.}

The idea of Drive-thru Internet by connecting to existing roadside WiFi
Access Points is introduced in~\cite{drive-thru}, which shows that a
single moving vehicle connected via 802.11b with an AP located at
roadside of an empty street can access several megabytes of TCP or
UDP traffic, even when the velocity is as high as 180 km/h.
Subsequently, evaluations in various controlled
environments~\cite{drive-thru,in-motion,MobiSteer,Interactive} and
in situ WiFi networks~\cite{CarTel,MobiSteer,Interactive,Cabernet}
have been conducted, further confirming the feasibility of
WiFi-based Vehicular Internet Access for non-interactive
applications. More recently, the performance of content delivery using unicast and broadcast from road-side WiFi infostations is studied in~\cite{ZhengHaiTao-RoadsideCDS}.
In addition to WiFi, small cell architectures such as femtocells, which were initially designed to improve the indoor cellular experience, are being extended to provide high data rate coverage over short ranges to the outdoor environment~\cite{QualcommSmallCell,UbiquisysSmallCell}.

In spite of these efforts, scalable solutions for the {\it deployment} and {\it management} of WiFi APs or femtocell base-stations to enable efficient vehicular Internet Access have not been fully understood so far. Instead, simple heuristics without performance guarantees are
commonly adopted in most previous works. For instance, a simple non-uniform
strategy that places more stationary nodes in the network core was
considered in a recent work~\cite{Relay}. In addition to initial or incremental deployment at a relatively large time scale, optimal network design is also important for dynamic network management to sustain an economical infrastructure or improve the revenue of the service provider. For instance, dynamic sleep-wakeup scheduling of base-stations has been viewed as a promising technique for improving the energy efficiency of the cellular infrastructure~\cite{traffic-driven}. However, current solutions are again based on simple heuristics without providing guaranteed quality of service to the end users~\cite{traffic-driven,Tassiulas-greenbs}.

Our previous work on Alpha Coverage~\cite{alphacover,alphacover-journal} initiated research on scalable deployment of road-side APs for providing guaranteed service to mobile vehicles. In Alpha Coverage, each AP is viewed as a point in the road network graph, and the objective is to bound the gap between two consecutive contacts, while ignoring the quality of each contact. In contrast, the notion of Contact Opportunity~\cite{contactopp} provides a more accurate and practical measurement of service quality for mobile entities by properly modeling the expected data rate from each contact and various uncertainties involved in the system.

Following our studies in~\cite{alphacover,alphacover-journal,contactopp}, a few works have also considered planned AP deployment for serving mobile vehicles~\cite{Malandrino-Content-TMC,Delay-Constrained-Coverage,RSU-game}. By assuming full knowledge of vehicle mobility in both space and time, the AP deployment problem for maximizing the total content downloading rate is formulated as an MILP problem in~\cite{Malandrino-Content-TMC}, where the possibility of using other vehicles as relays (through V2V communication) is also considered. Due to the high complexity of solving the MILP for a large system, a sampling-based solution is then proposed, which, however, does not provide any performance bound. In~\cite{Delay-Constrained-Coverage}, the AP deployment problem for collecting delay constrained sensing data generated from onboard sensing devices in vehicles is considered. By ignoring the capacity constraint of APs and focusing on a single (expected) scenario, the optimization problem reduces to the submodular maximization problem with a linear constraint, and the standard greedy algorithm can be applied. Furthermore, these works ignore the uncertainties associated with the road traffic and the available data rate of APs. In~\cite{RSU-game}, a non-cooperative game for road-side unit (RSU) deployment in the context of multiple operators is considered under a simplified setting with a single road segment and two candidate locations.

Submodular functions play a critical role in combinatorial
optimization. The theory was first developed half a century ago.
Since then, various submodular optimization problems have been
intensively studied~\cite{combinatorial-optimization, submodular}.
Although the submodular minimization problem is polynomial time
solvable, the submodular maximization problem is NP-hard. However, a
simple greedy algorithm gives an $(1-1/e)$ approximation ratio when
the function is also nondecreasing and normalized. A similar greedy solution also applies to the submodular
set covering problem first studied in~\cite{submodularsetcover} and achieves a logarithmic factor.
Recently, submodular optimization has been applied to several
network deployment problems in the context of sensor networks, including placing sensors to
efficiently detect outbreak~\cite{outbreak}, or to provide robust
observations~\cite{robust}. The former is formulated as a submodular
maximization problem while the latter is formulated as a budgeted
submodular set covering problem. On the other hand, submodular covering in the face of uncertainty has only been
studied in some special cases, e.g., set cover and vertex cover,
for both robust~\cite{Feige-robust} and stochastic optimization~\cite{Shmoys-twostage,Shmoys-twostage-survey}. In this work, we show that the greedy algorithm for submodular covering can be extended to the polynomial-scenario setting, which can then be combined with the sample average approximation method~\cite{Shmoys-Shmoys-SAA} to obtain an efficient solution to our deployment problem under the two-stage stochastic setting.

\section{Contact Opportunity Optimization}\label{sec:acquire}
Ideally, we would like to have a scalable deployment of APs that is
able to serve mobile users on the go with
guaranteed performance in terms of some intuitive metric such as
\ignore{long term} average throughput. Such an objective is
complicated by various uncertainties in the system, such as
unpredictable traffic conditions, unknown moving patterns of mobile
users, and the dynamics involved in the performance of APs. To this
end, we use an incremental approach; we introduce a performance
metric for roadside AP deployment that is closely related to average
throughput while avoiding the uncertainties such that an efficient
solution can be obtained.
In Section~\ref{sec:extension}, several extensions that consider
more intuitive performance metrics and more practical system models
are introduced.

\subsection{System Model}
We model a road network as a connected {\it geometric} graph, where
vertices represent points where road centerline segments and road
intersections meet, and edges represent road centerline segments
connecting road intersections. For a curved road segment, we
introduce artificial road intersections, so that each edge
represents a straight line segment. Without loss of generality, the
road network graph is assumed to be undirected. Let $d_e$ denote the length of road segment $e$, and let $E$ denote the set of road segments. \ignore{The problem
definitions and solution presented in this paper can be applied to
directed graphs without modification.}

\begin{figure}
\centering
\includegraphics[width=1.8in]{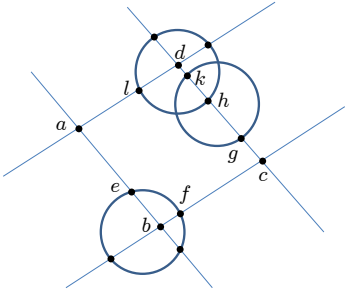}
\caption{\small A road network with four roads (lines) and three
candidate locations with coverage regions shown as disks. There are four road intersections, i.e., $a$, $b$, $c$, and $d$.
The coverage disks partition the roads into subsegments such as $ae,be,bf,cg,dl$, etc.}
\label{fig:seg}
\end{figure}

In this section, we focus on the deployment of new APs that serve mobile users exclusively.
Extensions to acquiring service from existing APs and the coexistence of static and mobile users will be considered in Section~\ref{sec:extension}.
Let $A$ denote a set of known candidate locations in the 2D
region covering the road network where new APs can be deployed. Note that any points in the 2D area can be a candidate location, although for simplicity,
we take the set of road intersections as candidate locations in our simulations.
Associated with each candidate location $a \in A$, there is a fixed
cost $w_a \in \mathbb{R}^+$ for installing an AP at $a$, and a
coverage region $C_a$, which is a connected region in the 2D space
consisting of the set of points where the received SNR from an AP
deployed at $a$ is higher than a fixed threshold. The coverage
regions partition the road network graph into
smaller segments called {\it subsegments}.
Figure~\ref{fig:seg} shows a road network with four roads (lines)
and three candidate locations with coverage regions shown as disks, which partition the roads into subsegments such as $ae,be,bf,cg,dl$,
etc. A subsegment may be covered by multiple coverage regions, such as $hk$, or not covered at all, such as $al$. Although the coverage regions are plotted as disks in
Figure~\ref{fig:seg}, our problem definitions and solutions are
independent of the shape of a coverage region.

Let $L$ denote the set of all the subsegments in the road
network graph with respect to $A$. For each $l \in L$, let $d_l \in
\mathbb{R}^+$ denote the length of the corresponding road centerline
segment. Let $L_e \subseteq L$ denote the set of subsegments on edge $e \in E$.
For any deployment $S \subseteq A$, let $L_S \subseteq L$ denote the set of subsegments covered by $S$,
that is, $L_S = \{l \in L: l \subseteq
\cup_{a \in S}C_a \}$. 

A movement on a road network is modeled as a simple path on the
corresponding graph. We assume that there is a set of movements,
denoted as $P$, given as part of the input to the
deployment decision maker. For
instance, $P$ could be a set of shortest (or fastest) paths
or a set of most frequently traveled paths between a set of sources and destinations.
Such information can be learned from a road network database~\cite{Tiger} and historical
traffic data~\cite{vldb07}. The concrete definition of $P$
is independent of our problem definitions and solutions, while the
size of the set $P$ impacts the computational complexity
and performance guarantee of our solutions as discussed below. For each $p \in P$, let $E_p \subseteq E$ denote the set of edges on $p$, and
$L_p \subseteq L$ the set of subsegments on $p$.

\subsection{Problem Statement}
We now define a performance metric for roadside deployment that does
not require any information about the dynamics of the system. Given
a deployment $S \subseteq A$, the {\bf Contact Opportunity in
Distance} of a path $p \in P$, denoted as $\eta^d_p$, is
defined as the fraction of distance on $p$ that is covered by some
AP in $S$. Formally,
\begin{equation} \eta^d_p (S) = \frac{\sum_{l \in L_p \cap L_S} d_l}{\sum_{l \in L_p} d_l}.
\end{equation}

\begin{table}
\renewcommand{\arraystretch}{1.3}
\caption{\small Notation List}
\label{table_example} \centering
\begin{tabular}{|l|l|}
\hline $E$ & Set of road segments  \\
\hline $V$ & Set of road intersections  \\
\hline $d_e$ & Length of road segment $e$\\
\hline $v_e$ & Driving speed on road segment $e$\\
\hline $h_e$ & Traffic density on road segment $e$\\
\hline $A$ & Set of all candidate locations for deploying APs \\
\hline $w_a$ & Cost of installing an AP at location $a \in A$ \\
\hline $C_a$ & Coverage region of an AP deployed at $a \in A$ \\
\hline $r_a$ & Available data rate of an AP at $a$ for mobile users\\
\hline $L$ & Set of road subsegments\\
\hline $d_l$ & Length of a subsegment $l \in L$ \\
\hline $n_l$ & Number of mobile users on a subsegment $l$ \\
\hline $r_l$ & Data rate for serving a mobile user on a subsegment $l$ \\
\hline $L_S$ & Set of subsegments covered by $S \subseteq A$\\
\hline $L_e$ & Set of subsegments that belong to segment $e \in E$\\
\hline $P$ & Set of movements (paths)\\
\hline $E_p$ & Set of road segments on path $p$\\
\hline $L_p$ & Set of subsegments on path $p$\\
\hline $A_p$ & Set of candidate locations that cover path $p$\\
\hline $\eta^d_p$ & Contact opportunity in distance over path $p$\\
\hline $\eta^t_p$ & Contact opportunity in time over path $p$\\
\hline $\gamma_p$ & Average throughout over a path $p$\\
\hline $K$  & Set of possible scenarios for travel speed, user density, and traffic load\\
\hline $k_S$ & A worst-case scenario in $K$ with respect to a deployment $S$\\
\hline
\end{tabular}
\end{table}

When a mobile user travels at a constant speed where each AP has the
same data rate, and there is only one user in the system, contact
opportunity in distance can be directly translated into average
throughput that the user will experience. We show in
Section~\ref{sec:extension} how to extend this concept by taking
various dynamic elements into account. \ignore{Given a budget $B$, we are looking
for a deployment where the total cost of the APs deployed does not
exceed the budget, and the minimum contact opportunity over all
movements in $P$ is maximized. Such a deployment provides a
worst case guarantee and hence does not require statistics about the
movement patterns of mobile users. Furthermore, simulation results
(in Section~\ref{sec:sim}) show that even though our solution is
designed for the worst case, it works well in the average case.}
Our objective is to provide a required level of contact opportunity over all the
movements in $P$ at a minimum cost.
Formally, let $\lambda_p$ denote the required contact opportunity for path $p$, and $w(S)$ the cost of a deployment $S
\subseteq A$, that is, $w(S) = \sum_{a \in S}w_a$, the first optimization
problem that we consider is:
\begin{equation}\label{prob:budget}\min_{S \subseteq A} w(S), \textrm{ subject to } \eta^d_p (S) \geq \lambda_p, \forall p \in P.
\end{equation}

For a given set of parameters $\lambda_p$, let $\lambda = \min_{p \in P} \lambda_p$ and $\tilde{\eta}^d_p(S) = \eta^d_p(S) \times \lambda /\lambda_p$.
Then the constraint in (2) is equivalent to requiring that the minimum $\tilde{\eta}^d_p(S)$ among {\it all} the paths in $P$ is at least $\lambda$. To simplify the notation, we will use $\eta^d_p$ to denote $\tilde{\eta}^d_p$
in the rest of the paper. Thus, the optimization problem to be solved is:
\begin{align}
\textbf{ P1:} \hspace{2ex} & \min_{S \subseteq A} \hspace{1ex} w(S) \nonumber\\
& \textrm{s.t.} \hspace{1ex} \min_{p \in P} \eta^d_p(S) \geq \lambda
\end{align}
\vspace{-5pt}

We will also study a dual problem that maximizes the minimum contact opportunity among all the paths for a given budget $B$:
\begin{align}
\textbf{ P2:} \hspace{2ex} & \max_{S \subseteq A} \min_{p \in P} \hspace{1ex} \eta^d_p(S) \nonumber\\
& \textrm{s.t.} \hspace{1ex} w(S) \leq B
\end{align}
\vspace{-5pt}

\noindent {\bf Hardness of the problem:} We note that both problems are NP-hard in general. To see this, consider a road network graph where each vertex is a candidate location for APs. Assume that the coverage region of an AP at vertex $a$ can fully cover all the edges incident to $a$ and only those edges, and the set of movements in $P$ are paths consisting of single edges. Then a reduction from Vertex Cover to the decision version of our problems can be easily constructed. Since Vertex Cover is NP-complete even when restricted to 3-connected,
cubic planar graphs~\cite{Uehara}, both \textbf{P1} and \textbf{P2} are NP-hard. Hence, it is not likely that optimal solutions to these problems can be obtained for most practical settings.
Our approach is to design efficient approximation algorithms that can be implemented even in a large scale system,
while ensuring a guaranteed performance.

\subsection{Minimum Cost Contact Opportunity} \label{sec:submodular}
In this section, we first present a simple greedy algorithm to \textbf{P1} and show that the algorithm achieves a guaranteed performance, by a reduction to the submodular set covering problem~\cite{submodularsetcover}. We then discuss strategies to accelerate the computation in our context.

\vspace{1ex}
\noindent {\bf A Greedy Algorithm:} We first note that if we define
\begin{equation}\label{def:trunk}
\eta^d(S,\lambda) = \sum_{p \in P}\min \{\eta^d_p(S), \lambda\},
\end{equation}

\noindent then a subset $S \subseteq A$ is a feasible solution to \textbf{P1}
iff $\eta^d(S, \lambda) = \eta^d(A,\lambda) = \lambda|P|$. To see this, first note that $S$ is feasible iff $\eta^d_p(S) \geq \lambda$ for all $p$ by the problem definition, which is true iff $\eta^d(S, \lambda) = \lambda|P|$ by~\eqref{def:trunk}. Moreover, \textbf{P1} has a feasible solution iff $A$ is feasible. Hence the statement holds. Based on this observation,
the greedy algorithm for \textbf{P1} is sketched in Algorithm~\ref{alg:bic}.
The algorithm starts with an empty set and in each iteration picks a new candidate location that is most cost-effective, i.e., the location that maximizes the incremental difference (normalized by the weight). The procedure repeats until the
required contact opportunity is achieved.

\begin{algorithm}
\caption{Minimum Cost Contact Opportunity}\label{alg:bic}{\small
Input: $A, P, \lambda$\\
Output: A subset $S \subseteq A$}
\begin{algorithmic}[1]
\State $S \leftarrow \emptyset$; \While{$\eta^d(S,\lambda) <
\eta^d(A,\lambda)$}\State Find $a \in A \backslash S$ that maximizes
$\frac{\eta^d(S \cup \{a\},\lambda) - \eta^d(S,\lambda)}{w_a}$; \State $S \leftarrow
S \cup \{a\}$; \EndWhile
\end{algorithmic}
\end{algorithm}

\vspace{1ex}
\noindent {\bf Approximation Analysis:} To prove an approximation factor to Algorithm~\ref{alg:bic}, we first observe some structural properties of $\eta^d_p(S)$ and $\eta^d(S,\lambda)$.
In particular, we note that the set function $\eta^d_p: 2^A \rightarrow [0,1]$ satisfies the following
properties: (1) nondecreasing, i.e., $\eta^d_p(S) \leq \eta^d_p(T)$
whenever $S \subseteq T \subseteq A$; (2) normalized, i.e.,
$\eta^d_p(\emptyset) = 0$; and (3) \emph{submodular}, i.e., for all
$S \subseteq T \subseteq A$ and $a \in A \backslash T$, $\eta^d_p(S
\cup \{a\}) - \eta^d_p(S) \geq \eta^d_p(T \cup \{a\}) -
\eta^d_p(T)$. The last property is formally proved below, which
essentially says that \emph{adding a new AP to a small set helps
more than adding it to a large set}. It captures our intuition that
the total coverage that two APs can provide to a mobile user is
reduced if their communication regions overlap with each other.

\begin{lemma}\label{lemma:submodular}
$\eta^d_p(\cdot)$ is submodular.
\end{lemma}
\begin{IEEEproof}
For any $S \subseteq T \subseteq A, a \in A \backslash T$,
$\eta^d_p(S \cup \{a\}) - \eta^d_p(S) = \frac{\sum_{l \in
L_p \cap (L_{\{a\}} \backslash L_S)}
d_l}{\sum_{l \in L_p} d_l}$, and $\eta^d_p(T \cup \{a\}) -
\eta^d_p(T) = \frac{\sum_{l \in L_p \cap
(L_{\{a\}} \backslash L_T)} d_l}{\sum_{l \in
L_p} d_l}$. Since $S \subseteq T$, $L_{\{a\}}
\backslash L_S \supseteq L_{\{a\}} \backslash
L_T$. Therefore, $\eta^d_p(S \cup \{a\}) - \eta^d_p(S)
\geq \eta^d_p(T \cup \{a\}) - \eta^d_p(T)$.
\end{IEEEproof}

We then note that $\eta^d(\cdot,\lambda)$ for a given $\lambda$ is also a monotone submodular function since (a) $\min
\{\eta^d_p(S), \lambda\}$ as a set function over subsets of $A$ is submodular
when $\eta^d_p$ is submodular~\cite{combinatorial-optimization} and
(b) the sum of submodular functions is submodular.

It follows that \textbf{P1} is an instance of the submodular set covering
problem~\cite{submodularsetcover,robust}. In the general form of the problem, we are given a submodular function $f(\cdot)$ defined on a set $A$, and a cost $w(a)$ for any $a \in A$, the objective is to find a subset $S$ to minimize $w(S)$ such that $f(S) \geq \lambda$. We then have the following performance guarantee:

\begin{proposition}\label{theo:cost}
Algorithm~\ref{alg:bic} finds a feasible solution, the cost of which never exceeds the optimal cost by more than a factor $O(1)+\log(\max_{a \in A}D_a)$, where $D_a = \sum_{p \in P}\sum_{l \in L_p \cap L_{\{a\}}}d_l$ denotes the total distance covered by a single AP $a \in A$ over all the paths
\end{proposition}
\begin{IEEEproof}
A classical result in~\cite{submodularsetcover} is that when $f$ is monotone submodular and has {\it integer} values, the greedy algorithm has an approximation factor of $O(1)+ \log (\max_{a \in A} f(\{a\}))$.
To apply this result in our context, we rewrite the constraint (3) in \textbf{P1} as $\sum_{l \in L_p \cap L_S} d_l \leq \lambda \sum_{l \in L_p} d_l$ for each $p$. By taking a proper unit, we can assume all the distance values are integral. For a given $\lambda$, if we define $f(S) = \sum_p \min\{\sum_{l \in L_p \cap L_S} d_l, \lambda \sum_{l \in L_p} d_l\}$, our problem becomes a submodular set covering problem with respect to $f$. Hence, we get an approximation factor of $O(1) +
\log (\max_a \min\{\sum_{l \in L_p \cap L_{\{a\}}} d_l, \lambda \sum_{l \in L_p} d_l\}) \leq O(1) +  \log (\max_a \sum_{l \in L_p \cap L_{\{a\}}} d_l) = O(1)+\log(\max_a D_a)$
\end{IEEEproof}

The above procedure can be naturally extended to improving an existing deployment by adding new APs, by substituting
all the evaluations of $\eta^d_p(S)$ with $\eta^d_p(S \cup A_0)$,
where $A_0$ indicates the set of APs previously deployed. \ignore{The
same performance guarantee applies to this incremental deployment as
well.}

\vspace{1ex}
\noindent{\bf Techniques to Accelerate the Computation}:
Algorithm~\ref{alg:bic} requires O($|A|$) iterations (line 2 to line
4) where each iteration involves $|A|$ evaluations of $\eta^d(\cdot, \lambda)$.
Hence, in all it requires $O(|A|^2)$ evaluations of $\eta^d(\cdot, \lambda)$, where each
evaluation involves computing $\eta^d_p(\cdot, \lambda)$ for each $p \in
P$, which takes $O(|P||V||A|)$ time. Hence, the total complexity is $O(|P||V||A|^3)$, which is very time consuming for a large road network,
and with large $|A|$ and $|P|$. Below we propose several
techniques to accelerate the computation in our context.

\begin{itemize}
\item First, we apply the accelerated greedy algorithm~\cite{AG} to our problem, which significantly reduces the total number of evaluations of $\eta^d(\cdot, \lambda)$ needed through lazy evaluations. The submodularity of $\eta^d(\cdot,\lambda)$ implies that the incremental difference $\eta^d(S \cup \{a\},\lambda) - \eta^d(S,\lambda)$ for any candidate location $a$ is non-increasing in $S$. In the algorithm, a priority queue is used to maintain a set of incremental differences for the candidate locations. In each iteration, instead of checking all the possible candidate locations as in the simple greedy algorithm, locations with higher incremental differences up to this stage are first considered, which avoids a large number of evaluations. More details can be found in~\cite{AG,accelerated-greedy-d-optimal}.

\item Second, we note that for any path $p \in P$, if $p$ can be
divided at certain road intersection into two sub-paths $p_1,p_2
\in P$ such that $\eta^d_{p_1}(S) \geq \lambda,
\eta^d_{p_2}(S) \geq \lambda$, then $\eta^d_p(S) \geq \lambda$ as
well. In this case, one can safely exclude $p$ from $P$ without loss of optimality. Now suppose $P$ is the composed of all the shortest
paths of length at least $\alpha$ in the road network graph, and the maximum edge length
in the graph is significantly less than $\alpha$. Then it is highly likely that a shortest path
of length greater than $2\alpha$ can be divided into sub-paths
of length between $\alpha$ and $2\alpha$. These longer paths can then be dropped to reduce the size of $P$, which helps with both the time complexity of the algorithm and its performance guarantee. 

\item Third, we note that each candidate location only contributes to a small
subset of $\hat{P}$, and therefore an incremental
calculation is more efficient, where $\eta^d(S \cup \{a\},\lambda)$ is
obtained from $\eta^d(S,\lambda)$ by updating only $\eta^d_p$ for those $p$
covered by $C_a$.
\end{itemize}

We observe that these techniques improve the performance of our
algorithm significantly in practice. For the $6 \times 6$ km$^2$ road
network and a set of 10000 movements considered in our simulations (Section~\ref{sec:sim}), the
running time of Algorithm~\ref{alg:bic} to find a solution to \textbf{P1} is reduced
from hours to a few seconds under the same machine
configuration.

\subsection{Contact Opportunity Maximization}
After providing an approximation algorithm to Problem \textbf{P1}, we now propose a solution to \textbf{P2} by utilizing Algorithm~\ref{alg:bic} as a subroutine.
The idea is to apply a binary search over $\lambda \in [0,1]$. A similar approach has been applied in~\cite{robust} to solve a submodular covering problem with a budget.
The procedure is sketched in Algorithm~\ref{alg:search}.

\begin{algorithm}
\caption{Maximum Contact Opportunity}\label{alg:search}{\small
Input: $A, P, B$\\
Output: A subset $S \subseteq A$}
\begin{algorithmic}[1]
\State $\lambda_1 \leftarrow \min_{p \in P}\eta^d_p(A)$; $\lambda_2 = 0$;
\While{$\lambda_1 - \lambda_2 \geq \delta$}
\State $\lambda = (\lambda_1 + \lambda_2)/2$;
\State $S \leftarrow$ call Algorithm~\ref{alg:bic} with parameters $A, P$, $\lambda$;
\If{$w(S)>B$}
\State $\lambda_1 \leftarrow \lambda$;
\Else \hspace{1ex} $\lambda_2 \leftarrow \lambda$
\EndIf
\EndWhile
\end{algorithmic}
\end{algorithm}

The algorithm maintains an upper bound and a lower bound for achievable $\lambda$, denoted as $\lambda_1$ and $\lambda_2$, respectively.
Initially, $\lambda_1 = \min_{p \in
P}\eta^d_p(A)$, the minimum contact opportunity that can be achieved
when all the candidate locations are utilized to deploy APs, and $\lambda_2 = 0$. Algorithm~\ref{alg:bic} is then invoked with $\lambda = (\lambda_1 + \lambda_2)/2$ as the input.
If the solution found surpasses the budget, the upper bound is decreased; otherwise, the lower bound is increased. The procedure continues until the difference between the upper and lower bounds is
less than $\delta$, where $\delta$ can be adjusted to trade accuracy with computational time. Note that to accelerate the computation, an extra condition can be added to the {\textsf while} loop of Algorithm~\ref{alg:bic} (line 2) so that whenever
the current $S$ maintained already
violates the budget constraint, the above procedure can move on to a new $\lambda$.

For a given a budget $B$, the above binary search procedure always
finds a feasible deployment. Moreover, the algorithm achieves a bi-criteria approximation as stated below.


\begin{proposition}\label{prop:max-cover}
Given a budget $B$, let $\lambda(B)$ and $\lambda^*(B)$ denote the contact opportunity achieved by Algorithm~\ref{alg:search} and the optimal solution, respectively.
Then we have $\lambda(B) \geq \lambda^*(B/\epsilon)-\delta$, where $\epsilon = O(1)+\ln(\max_{a \in A}D_a)$.
\end{proposition}
\begin{IEEEproof}
In the binary search, if the value of $\lambda$ is set to $\lambda^*(B/\epsilon)$, then Algorithm~\ref{alg:bic} with this $\lambda$ as input will find a deployment $S$ of cost at most $B$
according to Proposition~\ref{theo:cost}. Since $S$ is feasible, the binary search won't miss this $\lambda$ beyond the small gap defined by $\delta$.
\end{IEEEproof}

\section{From Contact Opportunity to Average Throughput \label{sec:extension}}
The concept of contact opportunity in distance discussed in Section~\ref{sec:acquire}
ignores several complexities involved in a real system and does not
directly correspond to the quality of service for mobile users driving through the system. Therefore, we seek to
design performance metrics that are more intuitive to mobile
application designers and end users. In this section, we first extend the notion of contact opportunity to average throughput by modeling various
uncertainties involved in the system (Section~\ref{sec:uncertainty}). We then study the deployment problem of achieving a required level of average throughput while minimizing the worst-case cost or the expected cost. To this end, we will consider a robust optimization approach in Section~\ref{sec:robust}, and a two-stage stochastic optimization approach in Section~\ref{sec:two-stage}, respectively.

\subsection{Modeling Average Throughput Under Uncertainty}\label{sec:uncertainty}
To obtain a meaningful definition of average throughput in our context,
we start with modeling two key dynamic aspects in our system: road traffic conditions and the data traffic from static users.

First, it is clear that the average throughput that a mobile user can obtain depends on both its travel speed and the contact duration when it is associated with some APs.
However, both the contact time and the travel time are not fixed due to
the uncertainties of traffic conditions such as traffic jams,
accidents and stop signs. Moreover, the traffic condition also affects the number of mobile users that are in the range of the same AP at the same time competing for the bandwidth of the AP.
To model these uncertainties, we follow
the interval based modeling approach from~\cite{robust-discrete} and consider two key parameters in characterizing a traffic flow: speed and density~\cite{traffic-flow}.
More concretely, we assume that for each road segment $e \in E$, the {\it driving speed} on $e$, denoted as
$v_e$, is most likely within an interval $[v^1_e,v^2_e]$ for some constants $v^1_e>0$, $v^2_e>0$ and $v^1_e \leq
v^2_e$. Note that it is reasonable to assume a constant driving speed for a road segment. Similarly, the {\it traffic density} on road segment $e$, i.e., the number of road-side WiFi service users on $e$ per unit distance, denoted as $h_e$, is most likely within an interval $[h^1_e,h^2_e]$ where $0 < h^1_e \leq h^2_e$.

Second, as stated before, we envision a deployment model where both new APs can be installed and existing APs targeting at static users can be acquired to share their service with mobile users at certain cost.  With a slight abuse of notation, we again let $A$ denote the (disjoint) union of both candidate locations for new APs and the locations of existing APs that can be utilized. Our objective is to guarantee the service quality to mobile users at the minimum cost without affecting static users. Since the data traffic of static users often vary over time, the available date rate of an AP located at $a$ for serving mobile users, denoted by $r_a$, is a random variable. We again use the interval based approach and assume that the value of $r_a$ is most likely within an interval $[r^1_a, r^2_a]$, where $0 < r^1_a \leq r^2_a$.

The above intervals for modeling uncertainties are assumed to be given or can be learned from historical data. 
We define a \emph{scenario} $k$ to be an assignment of values to the random variables defined above from the corresponding intervals. Let $v_e(k), h_e(k)$ and $r_a(k)$ denote the corresponding values of these variables in a scenario $k$. We define $v_l(k)$ and $h_l(k)$ as the speed and density for a subsegment $l$, derived from the corresponding values of the road segment that $l$ belongs to. Let $K$ denote the set of
all possible scenarios. Note that $K$ is an infinite set. We then state the first natural extension to the notion of contact opportunity in distance, where we replace distance with time. Formally, given a deployment $S \subseteq A$ and a scenario $k \in K$, we define the {\bf Contact Opportunity in Time} of a path $p \in P$ as:
\begin{equation}\eta^{t}_p(S,k) = \frac{\sum_{l \in L_p \cap L_S} d_l/v_l(k)}{\sum_{l \in L_p} d_l/v_l(k)},
\end{equation}

\noindent which captures the fraction of {\it time} that a mobile user is in
contact with some AP when moving through $p$.

To move one step further and model average throughput, we need to make further assumptions regarding association control and scheduling in serving mobile traffic load. Consider a scenario $k$ and a deployment $S$. Let $u_l(k)$ denote the expected number of mobile users on a subsegment $l$, which can be estimated as $u_l(k) = h_l(k) d_l$. 
We will focus on the steady state where the mobile users in the system are distributed according to above estimates. 
In theory, an optimal scheduling policy that maximizes the time average throughput over a movement can be derived by solving a maximum flow problem. However, due to its high complexity and the centralized nature, such a policy is not likely to be used for serving real time traffic. Instead, we focus on simple stateless and distributed strategies that are easily implementable. Our approach is to estimate the expected data rate that a mobile user on a subsegment $l$ can obtain in the steady state, denoted as $r_l(k)$. Given the estimates, the {\bf Average Throughput} for a mobile user moving through a path $p$, denoted as $\gamma_p(S,k)$, can be stated as:
\begin{equation}\gamma_p(S,k) = \frac{\sum_{l \in L_p \cap L_S} [d_l/v_l(k)]r_l(k)}{\sum_{l \in L_p} d_l/v_l(k)}.\label{equ:throughput}
\end{equation}

Below we outline one approach to estimate $r_l(k)$. We consider the simple association protocol where each mobile user picks an AP in its range at random to associate, while APs serve all the users associated with in an equal rate. This protocol does not rely on any real-time information and can be easily implemented in practice. However, our deployment algorithms can be applied to more sophisticated association protocols as well. An interesting open problem is the joint optimization of deployment and association control, which is part of our future work. Consider a subsegment $l$ that is within the coverage regions of multiple APs. These APs are assumed to operate on orthogonal channels, and do not interfere with each other. By the random association assumption, a user in $l$ has an equal chance to be served by any of these APs. Let $n_l$ denote the number of APs that cover $l$, then $u_l/n_l$ users are assigned to each of these APs. Now for any AP $a$, let $L_a$ denote the set of subsegments in its range, and let $u_a =  \sum_{l \in L_a} u_l/n_l$ denote the total number of users associated with $a$. We assume that the AP serves all these users in an equal rate of $r_a/u_a$. For any user on segment $l$, its expected data rate can then be estimated as $r_l = \frac{\sum_{a \in S_l} r_a/u_a}{|S_l|}$. We observe that under this approach, average throughput reduces to contact opportunity in time when $r_a = 1$ and $u_a = 1$ for all $a \in A$.

\ignore{First, consider a coverage region $C_a$ that is disjoint from other coverage regions. Without maintaining any service history and real time information, it is reasonable to assume that all the users in $C_a$ have an equal share of the data rate $r_a$. Hence, each of them obtains a data rate of $r_a/u_a$. Now consider a user on a subsegment $l$ that is within the coverage regions of multiple APs. These APs are assumed to operate on orthogonal channels, and hence do not interfere with each other. Let $S_l \subseteq S$ denote the set of APs in $S$ that cover $l$. If $u_l \geq |S_l|$, all the APs in $S_l$ can be utilized to serve different users on $l$. We then estimate the data rate for a user on segment $l$ to be $\hat{r}_l = \sum_{a \in S_l} r_a/u_a$. The intuition is that these users can share the total capacity of the APs in $S_l$, assuming that the connection setup time and the handoff time when a user switches between APs are small enough to be ignored. On the other hand, if $u_l < |S_l|$, not all the APs in $S_l$ can be used assuming each user has a single antenna. In this case, let $S'_l \subseteq S_l$ denote the $u_l$ APs with the highest $r_a/u_a$. We then estimate the data rate for a user on $l$ to be $\hat{r}_l = \sum_{a \in S'_l} r_a/u_a$.}

As in Problem {\bf P1}, our objective is to ensure the required average throughput at the minimum cost. Since the cost varies for different scenarios, we would like to minimize either the cost in the worst-case scenario, or the expected cost. We outline two approaches below.

\subsection{Robust Optimization}\label{sec:robust}
We first study a robust optimization approach. Although there are infinitely many scenarios, we seek to find a
deployment that performs well even in the worst case. To this end, we first present two problems to be studied, which extend {\bf P1} and {\bf P2}, respectively, with the objective of ensuring average throughput under a worst-case scenario. We then propose an efficient algorithm to identify a worst-case scenario for any given deployment, which is then utilized to derive our solutions to the robust optimization problems.

\vspace{1ex}
\noindent {\bf Problem Statement:}
Let $w_a$  denote either the cost for installing a new AP at location $a$ or the (one-time) cost to obtain service from an existing AP located at $a$, and again define $w(S) = \sum_{a \in S}w_a$. Our objective is to solve the following problem:
\begin{align}
\textbf{ P3:} \hspace{2ex} & \min_{S \subseteq A} \hspace{1ex} w(S) \nonumber\\
& \textrm{s.t.} \hspace{1ex} \min_{p \in P, k \in K} \gamma_p(S,k) \geq \lambda \label{C1}
\end{align}
\vspace{-5pt}

\noindent where~\eqref{C1} ensures that for a mobile user moving through any path in $P$, an average throughput of $\lambda$ is obtained under any scenario. We will also consider the dual problem:
\begin{align}
\textbf{ P4:} \hspace{2ex} & \max_{S \subseteq A} \min_{p \in P, k \in K} \hspace{1ex} \gamma_p(S,k) \nonumber\\
& \textrm{s.t.} \hspace{1ex} w(S) \leq B
\end{align}
\vspace{-5pt}

For a given deployment $S$, we define a scenario $k_S \in K$ to be a worst-case scenario if
$\min_{p \in P} \gamma_p(S,k)$ is minimized at $k_S$ among all the scenarios in $K$. Then the constraint~\eqref{C1} is equivalent to $\min_{p \in P}\gamma_p(S,k_S) \geq \lambda$. Note that there may exist multiple worst-case scenarios in general, since only the values associated with the road segments on and coverage regions touching the path with the minimum throughput matter.

\vspace{1ex}
\noindent {\bf Identifying a Worst-case Scenario:}
We now present an efficient algorithm to find a worst-case scenario, followed by our solutions to {\bf P3} and {\bf P4}. We first note that since $r_l$ only appears in the numerator of \eqref{equ:throughput}, $\gamma_p(S,k)$ is minimized by taking the minimum possible value of $r_l$, that is, by setting $r_a = r^1_a$ and $h_l = h^2_l$ for all $a \in A$ and $l \in L$. We again usee $r_l$ to denote this worst-case value when there is no confusion. We define $r_l = 0$ if $l \not \in L_S$. It remains to determine the values of $v_l$. Our intuition is that a worst-case scenario is most likely to happen when the traffic condition is such that the travel speed is slow on road segments with poor data access while the speed is high on road segments with good data access. Since a constant travel speed is assumed for each road segment, we can rewrite \eqref{equ:throughput} as follows, where we drop the index $k$ to simplify the notation:
\begin{equation}\gamma_p(S,\cdot)  = \frac{\sum_{e \in E_p} r_e(S)d_e/v_e}{\sum_{e \in E_p} d_e/v_e}.\label{equ:throughput2}
\end{equation}

\noindent where $r_e(S) = (\sum_{l \in L_e \cap L_S} d_l r_l)/d_e$ indicates the average data rate over $e$ under deployment $S$. The following proposition formalizes the above intuition (see the Appendix for a proof):
\begin{proposition}\label{prop:scenario}
For any path $p$, there is an assignment of $v_e$ for $e \in E_p$ that minimizes $\gamma_p$ such that the following two conditions are satisfied: (1) $v_e = v^1_e$ or $v_e = v^2_e$, for all $e \in E_p$; (2) there is an element $e^* \in E_p$, such that $v_e = v^1_e$ if $r_e \leq r_{e^*}$, and $v_e = v^2_e$ if $r_e > r_{e^*}$.
\end{proposition}

The proposition states that there is a worst-case scenario for a path $p$, where the driving speed of every edge in $p$ takes one of its boundary values, and moreover, the assignment satisfies a dichotomy condition according to their average data rate $r_e$. Based on this observation, an assignment of $v_e$ that achieves a worst-case scenario can be easily found by a search over all the edges in path $p$ to find the pivot $e^*$ that minimizes $\gamma_p$, as sketched in Algorithm~\ref{alg:search}. A worst-case scenario over all the paths can be then found by a search over $P$.

\begin{algorithm}
\caption{Worst-Case Scenario for a path $p$}\label{alg:scenario}{\small
Input: $S, p$; Output: $h_e, v_e, r_a$}
\begin{algorithmic}[1]
\State $h_e \leftarrow h^2_e, \forall e \in E_p$; $r_a \leftarrow r^1_a, \forall a \in S$;
\For{$e^* \in E_p$}
\For{$e \in E_p$}
\If{$r_e \leq r_{e^*}$}
\State $v_e = v^1_e$;
\Else \hspace{1ex} $v_e = v^2_e$;
\EndIf
\EndFor
\State $\gamma_p(e^*) =  \frac{\sum_{e \in E_p} r_ed_e/v_e}{\sum_{e \in p} d_e/v_e}$;
\EndFor
\State Output the scenario that gives the minimum $\gamma_p(e^*)$
\end{algorithmic}
\end{algorithm}

We remark that in the special case of contact opportunity in time, i.e., $r_l = 1$ for all $l \in L$, if we further allow the subsegments on the same edge to have different driving speeds, the worst-case scenario allows a simpler characterization as follows. Let $v_l \in [v^1_l, v^2_l]$ denote the possible speed on subsegment $l$. Then for a given deployment $S$, a worst-case scenario is obtained by setting $v_l = v^1_l$ if $l$ is not covered by $S$, and $v_l = v^2_l$ otherwise. To see this, note that the contact opportunity in time can be written as $\eta^t_p(S,\cdot) = \frac{t_1}{t_2+t_1}$ where $t_1$ denotes the travel time over the
set of subsegments in $p$ that are covered by $S$ and $t_2$ denotes
the travel time over the other subsegments in $p$. Hence
$\eta^{t}_p$ is minimized when $t_1$ is minimized and $t_2$ is
maximized, which happens under the above scenario.

\vspace{1ex}
\noindent {\bf Solutions to Robust Optimization:}
We then propose solutions to {\bf P3} and {\bf P4}. First note that, if we consider a fixed worst-case scenario for each deployment $S$, denoted as $k_S$,
$\gamma_p(S,k_S)$ can be viewed as a set function over $A$. Hence a natural first attempt to {\bf P3} is to apply Algorithm~\ref{alg:bic} by replacing $\eta^d_p(S)$ with $\gamma_p(S,k_S)$.
However, this approach does not provide a performance guarantee. The difficulty is that although for a given scenario $k$, $\gamma_p(S,k)$ is submodular by a similar argument as in Lemma~\ref{lemma:submodular}, $\gamma_p(S,k_S)$ is not submodular in general as stated in the following proposition (see the Appendix for a proof):

\begin{proposition}\label{prop:nonsubmodular}
$\gamma_p(S,k_S)$ as a set function over $A$ is nondecreasing and normalized, but
not submodular.
\end{proposition}

In fact, it has been observed that the robust versions of many optimization problems
are significantly more difficult than the original problems~\cite{robust-discrete}. Although an efficient solution
with guaranteed performance to the general problem remains open, we
propose the following two approaches as first steps that work well
in many practical cases.

Our {\bf first approach} applies when for every $p \in P$, the set of candidate
locations that cover $p$, denoted as $A_p$, has small cardinality.
The key idea is to view the constraint \eqref{C1} as requiring that an average throughput is guaranteed over all the paths {\it and} under all the scenarios. For any path $p$, to identify a worst-case scenario with respect to $p$, it suffices to only consider the worst-case scenarios with respect to subsets of $A_p$, namely, $\{k_S \in K: S \subseteq A_p\}$, since other APs do not affect the performance over $p$. Therefore, to identify a worst-case scenario overall all the paths, it suffices to consider $K' = \{k_S \in K: S \subseteq A_p \text{ for some } p \in P\}$. Note that the size of $K'$ is $\sum_{p \in P}2^{|A_p|}$, which is polynomial in $|A|$ and $|P|$ when $|A_p| = O(\log A)$ for any $p$.
We then define $\gamma(S,\lambda) = \sum_{p \in P, k \in K'}\min\{\gamma_p(S,k),\lambda\}$, which is again submodular. Algorithm~\ref{alg:bic} can then be applied to {\bf P3} by replacing $\eta^d(S,\lambda)$ with $\gamma(S,\lambda)$. This approach has polynomial time complexity when $|A_p| = O(\log A)$ for any $p$. Moreover, by a similar argument as in Proposition~\ref{theo:cost}, it achieves an approximation factor $O(1)+\log(\max_{a \in A}\mathcal{R}_a)$, where $\mathcal{R}_a = \sum_{p \in P, k \in K'} \sum_{e \in E_p} r_e(\{a\})d_e/v_e(k)$ indicates the total throughput contributed by a single AP $a \in A$ across all the paths and all the scenarios in $K'$.

Our {\bf second approach} is to approximate the deployment dependent worst-case scenario by
a single fixed scenario that is independent of the deployment chosen. Let $k_0$ denote the ``mean speed'' scenario with $v_e(k_0) =
(v^1_e+v^2_e)/2, h_e(k_0) = h^2_e, \forall e \in E$, and $r_a(k_0) = r^1_a, \forall a \in A$. It turns out that, if
$v^2_e/v^1_e$ is small for all $e \in E$, $k_0$ can be used as
a good approximation of the worst-case scenario. More concretely, we have the following proposition for any deployment $S$:

\begin{proposition}\label{theorem:robust}
If $v^2_e/v^1_e \leq \beta$ for all $e \in E$, then $\gamma_p(S,k_S) \leq
\gamma_p(S,k_0) \leq \beta \gamma_p(S,k_S)$ for any $p \in P$.
\end{proposition}

A formal proof is given in the Appendix. The proposition implies that if
$v^2_e/v^1_e$ is bounded above by a constant $\beta \geq 1$, then for any path $p$, the loss of average throughput by replacing the worst-case scenario with the ``mean speed" scenario is bounded by $\beta$. In fact, the second inequality holds between $k_0$ and any other scenario, not necessarily the worst-case scenario. Based on this observation, we then design an algorithm to {\bf P3} as sketched in Algorithm~\ref{alg:bic2}.

\begin{algorithm}
\caption{Robust Minimum Cost Contact Opportunity}\label{alg:bic2}{\small
Input: $A, P, \lambda$\\
Output: A subset $S \subseteq A$}
\begin{algorithmic}[1]
\State $v_e \leftarrow (v^1_e+v^2_e)/2, h_e \leftarrow h^1_e, \forall e \in E$; $r_a \leftarrow r^1_a, \forall a \in A$;
\State $m \leftarrow (\beta-1)/\tau$;
\For{$i =0$ to $m$}
\State $\lambda_0 \leftarrow (1+i\tau)\lambda$;
\State $S \leftarrow$ call Algorithm~\ref{alg:bic} with parameters $A, P$ and \\ \hspace{7.6ex} $\lambda_0$, where $\eta^d_p(S)$ is replaced by $\eta^d_p(S,k_0)$;
\If{$\min_{p \in P}\gamma_p(S,k_S) \geq \lambda$} break
\EndIf
\EndFor
\end{algorithmic}
\end{algorithm}

The algorithm searches over $\lambda_0 = \lambda, (1+\tau)\lambda, ..., \beta \lambda$, and for each $\lambda_0$, Algorithm~\ref{alg:bic} is invoked with $\eta^d_p(S)$ replaced by $\gamma_p(S,k_0)$.
The search repeats until a deployment that achieves an average throughput of at least $\lambda$ in the worst-case scenario is found. Note that such a deployment always exists, since by setting $\lambda_0 = \beta \lambda$, the deployment found by Algorithm~\ref{alg:bic} achieves an average throughput $\beta \lambda$ under scenario $k_0$, which ensures an average throughput of $\lambda$ in the worst-case scenario by Proposition~\ref{theorem:robust}. Furthermore, the algorithm achieves a bi-criteria approximation in the following sense: the cost of the solution found is no larger than $c^*(\beta\lambda)(O(1)+\log(\max_{a \in A}R_a))$, where $c^*(\beta\lambda)$ is the optimal cost for achieving an average throughput of $\beta\lambda$, and $R_a = \sum_{p \in P} \sum_{e \in E_p} r_e(\{a\},k_0)d_e/v_e(k_0)$ indicates the total throughput contributed by a single AP $a \in A$ across all the paths under the scenario $k_0$.

Proposition~\ref{theorem:robust} also leads to a simple solution to {\bf P4}. The idea is to simply invoke Algorithm~\ref{alg:search} for the ``mean speed'' scenario, that is, replacing $\eta^d_p(S)$ with $\gamma_p(S,k_0)$. This approach always gives a feasible solution, while the minimum average throughput across all the path achieved is at least $\frac{1}{\beta}\Big(\lambda^*(B/\epsilon')-\delta\Big)$, where $\epsilon' = O(1)+\log(\max_{a \in A}R_a)$, and $\lambda^*(B/\epsilon')$ is the optimal achievable value under the budget $B/\epsilon'$. Note that, compared with the non-robust version (Proposition~\ref{prop:max-cover}), an extra factor of $1/\beta$ is lost.

\subsection{Two-stage Stochastic Optimization}\label{sec:two-stage}
In contrast to robust optimization, our second approach to achieving an economical deployment under uncertainty focuses on minimizing the {\it expected} cost for ensuring a required level of average throughput, based on knowledge of the scenario distribution. In this section, we adopt the 2-stage stochastic approximation framework widely used in decision making under uncertainty~\cite{Shmoys-twostage-survey}, which has a natural interpretation in our context as discussed below. We propose an efficient approximation solution based on the sample average approximation (SAA) method~\cite{Shmoys-Shmoys-SAA}, combined with an extension of Algorithm~\ref{alg:bic}.

We envision a setting where a deployment is created in two stages, which can be readily generalized to the multi-stage case. In the first stage, the service provider implements an initial deployment by installing new APs or contracting with existing AP owners at selected locations. This decision is based on the prediction of system dynamics, such as road traffic condition and data traffic load from static users, for a relatively long period of time, say one month or one year. In the second stage, after the more accurate or actual traffic condition is realized, the initial deployment is augmented by acquiring service from additional APs, if needed, which happens at a relatively short time scale, say one day or one hour. Due to the short lead time in the second stage, it is expected that APs obtained in the second stage are more costly than that acquired in the first stage. Let $w^1_a$ denote the (amortized) cost per unit of time for an AP $a \in A$ installed/leased in the first stage, and $w^2_a>w^1_a$ the corresponding cost if it is acquired in the second stage. Let $w_1(S)=\sum_{a \in S}w^1_a$ and $w_2(S)=\sum_{a \in S}w^2_a$. Let $\mathcal{K}$ denote a random scenario with all the possible realizations in $K$. The two-stage optimization problem can be formulated as follows.
\begin{align}
\textbf{ P5:} \hspace{2ex} & \min_{S \subseteq A} \hspace{1ex} w_1(S) +  \mathbb{E}_{\mathcal{K}}(f_k(S))\nonumber\\
& \textrm{where} \hspace{1ex} f_k(S) = \min_{S_k \in A \backslash S} w_2(S_k) \nonumber \\
& \hspace{16.5ex} \textrm{s.t.} \hspace{2ex} \min_{p \in P} \gamma_p(S \cup S_k,k) \geq \lambda
\end{align}
\vspace{-5pt}

\noindent where the objective is to minimize the summation of the first stage cost and the expected second stage cost, with the expectation taken over all possible scenarios. For any scenario $k$ that is realized in the second stage, additional APs are deployed, if needed, with the objective of minimizing the second stage cost while ensuring a required average throughput under $k$. In general, both $w^2_a$ and $\lambda$ can depend on $k$. But we focus on the above problem for the sake of simplicity. A dual problem that maximizes the expected throughout subject to a budget on the total (two stage) cost can be similarly defined.

We emphasize that minimizing the expected cost is {\it different} from minimizing the cost of the expected scenario. The latter problem reduces to the single scenario case once the expected scenario is identified, and Algorithm~\ref{alg:bic} can be readily applied. On the other hand, minimizing the expected cost is significantly more difficult. It is known that some significantly simplified stochastic problems for minimizing the expected cost are \#P-hard even though their deterministic counterparts are polynomial time solvable~\cite{Shmoys-twostage}.

A fundamental challenge in {\bf P5} is due to the large number of possible scenarios, even if we discretize the scenarios and ignore the correlation in traffic distribution on nearby roads or APs. As a first step to address the challenge, we apply the sample average approximation method to reduce the infinite scenario problem to a polynomial-scenario problem. That is, a polynomial number of scenarios, denoted as $\mathcal{N}$, are first sampled by treating the distribution of scenarios as a black box, and then {\bf P5} is solved by considering only these scenarios. The objective function of {\bf P5} is replaced by $w_1(S) + \frac{1}{N}\sum_{k \in \mathcal{N}}f_k(S)$, where $N = |\mathcal{N}|$. It has been proved that for a large class of 2-stage stochastic linear programs, a polynomial number of samples is sufficient to ensure that an $\rho$-approximation solution to the sample-average problem is an $(\rho+\kappa)$-approximation solution to the original problem~\cite{Shmoys-Shmoys-SAA,Shmoys-twostage-survey} for some constant $\kappa>0$, where the polynomial bound depends on the input size, the maximum ratio between the second stage cost and the first stage cost, and $1/\kappa$. Although this bound cannot be directly applied to our problem, we expect that the SAA method provides a good performance for a reasonable number of samples, which is confirmed in simulations.

We then proceed to solve the polynomial-scenario problem, where we need to determine the initial deployment and the augmentation for each scenario in $\mathcal{N}$. As inspired by the stochastic set cover problem considered in~\cite{Ravi-stochastic}, we extend our definition of $\gamma_p(S,k)$ for the single-scenario case as follows. First, $N+1$ copies are created for each AP. Let $a^k$ denote the $k$-th copy of $a \in A$, with index $k \geq 1$ corresponds to the $k$-th scenario in $\mathcal{N}$, and index 0 corresponds to the initial deployment. The cost of $a^k$, denoted as $\tilde{w}_{a^k}$, is defined as $w^1_a$ if $k = 0$ and $\frac{1}{N}w^2_a$ if $k \geq 1$. Let $\mathcal{A}$ denote the set of all the copies of APs. Any subset $\mathcal{S} \subseteq \mathcal{A}$ then indicates a solution to the polynomial-scenario problem, with the initial deployment defined as $S_0 = \{a: a^0 \in \mathcal{S}\}$ and the augmentation in $k$-th scenario defined as $S_k = \{a: a^k \in \mathcal{S}\}$. The cost of a solution $\mathcal{S}$ is then defined as $\tilde{w}(\mathcal{S}) = \sum_{a^k \in \mathcal{S}}\tilde{w}_{a^k} = w_1(S_0)+\frac{1}{N}\sum^N_{k=1}w_2(S_k)$, which is the summation of the first-stage cost and the expected second stage cost respecting $\mathcal{S}$. For any scenario $k \in \mathcal{N}$, we define $\tilde{\gamma}_p(\mathcal{S},k) = \gamma_p(S_0 \cup S_k,k)$. The polynomial-scenario problem can then be refined as
\begin{align}
\textbf{ P6:} \hspace{2ex} & \min_{\mathcal{S} \subseteq \mathcal{A}} \hspace{1ex} \tilde{w}(\mathcal{S}) \nonumber\\
& \textrm{s.t.} \hspace{1ex} \min_{p \in P, k \in \mathcal{N}} \tilde{\gamma}_p(\mathcal{S},k) \geq \lambda
\end{align}
\vspace{-5pt}

Observe that {\bf P6} has a similar form to {\bf P3}. We then define $\tilde{\gamma}(\mathcal{S},\lambda) = \sum_{p \in P, k \in \mathcal{N}}\min(\tilde{\gamma}_p(\mathcal{S},k),\lambda)$, and observe that $\tilde{\gamma}(\mathcal{S},\lambda)$ is again monotone submodular. Hence, Algorithm~\ref{alg:bic} can be applied to {\bf P6} and achieves an approximation factor of $O(1)+\log(\max_{a \in A}\mathcal{R}'_a)$, where $\mathcal{R}'_a = \sum_{p \in P, k \in \mathcal{N}} \sum_{e \in E_p} r_e(\{a\},k)d_e/v_e(k)$ indicates the total throughput contributed by a single AP $a \in A$ across all the paths and all the scenarios in $\mathcal{N}$. Compared with the single scenario case, the approximation factor is worsen by an $O(\log N)$ factor. Therefore, although a larger $N$ improves sampling accuracy, it also incurs a worse approximation factor when solving the sampled problem. An interesting open problem is then to identify an optimal $N$ that balances the two effects.

The above solution has a complexity depending on $N$. We then consider a simple heuristic with a lower complexity. The idea is to find the initial deployment by simply applying Algorithm~\ref{alg:bic} to the ``mean'' scenario $k^0$, where $v_e(k^0) = (v^1_e+v^2_e)/2, h_e(k^0) = (h^1_e+h^2_e)/2, \forall e \in E$, and $r_a(k^0) = (r^1_a+r^2_a)/2, \forall a \in A$. Note the difference between $k^0$ and $k_0$ considered before. Also note that $k^0$ is the {\it expected} scenario when $v_e, h_e$, and $r_a$ are independently and uniformly distributed in the corresponding intervals. We will compare this heuristic and the SAA based approach in the simulations. 
\section{Simulations}\label{sec:sim}
In this section, we evaluate our roadside AP
deployment algorithms via numerical results and ns3-based simulations~\cite{ns3}, using real road networks
retrieved from 2008 Tiger/Line shapefiles~\cite{Tiger}. We compare Algorithms~\ref{alg:bic} and ~\ref{alg:search} and their extensions with two baseline algorithms to study the worst-case cost for achieving a required level of contact opportunity or average throughput under uncertainty, as well as the level of QoS guarantee that can be provided under a budget constraint. We further compare the SAA based algorithm with two heuristics to study the expected deployment cost under the two-stage setting.

\subsection{Numerical Results}
To understand the performance of our algorithms in a relatively large scale and under various parameter settings, we first resort to numerical study.

\begin{figure}[!ht]
\centerline{\subfigure{\includegraphics[width =
2in]{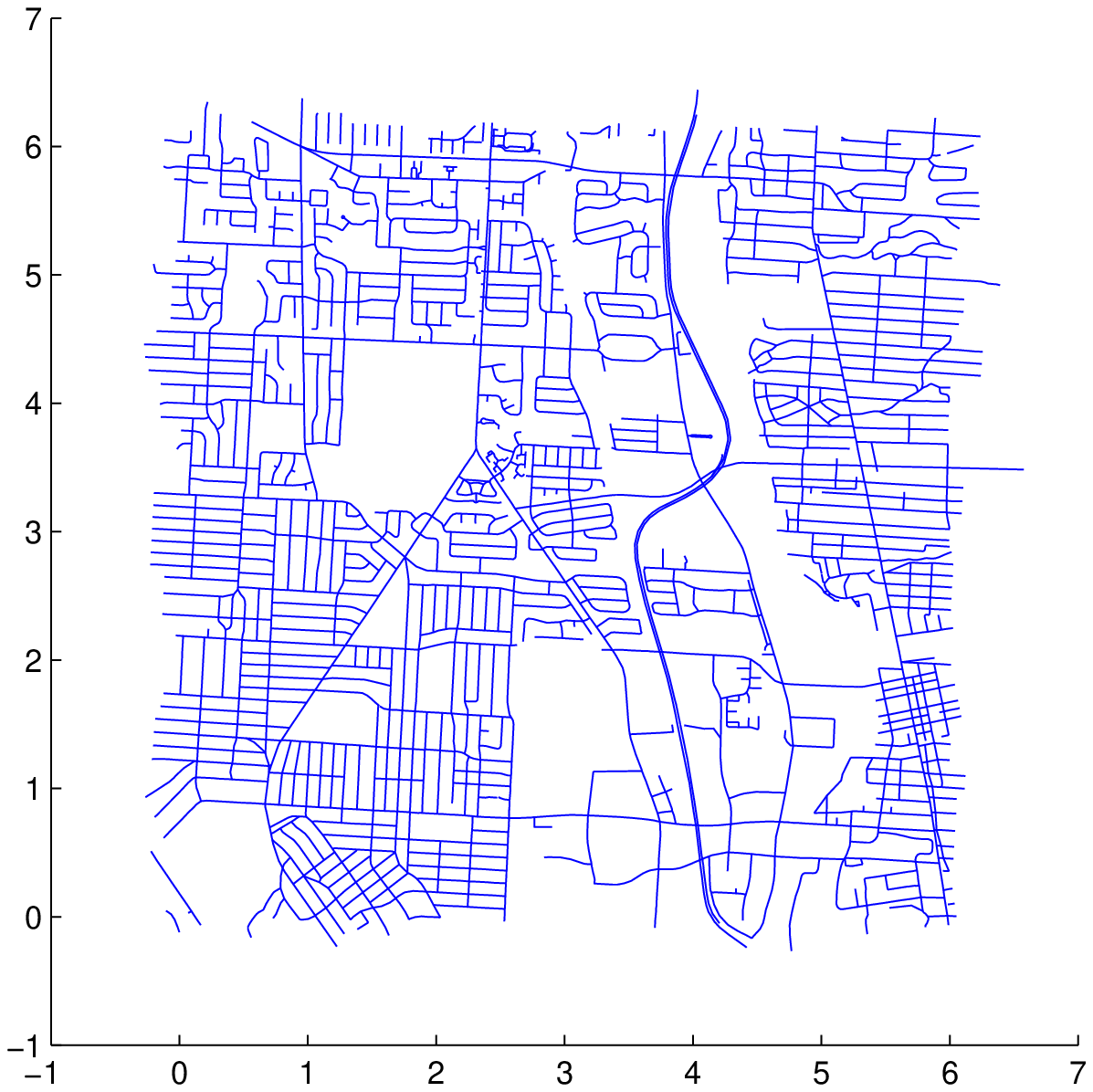}}
\subfigure{\includegraphics[width=1.2in]{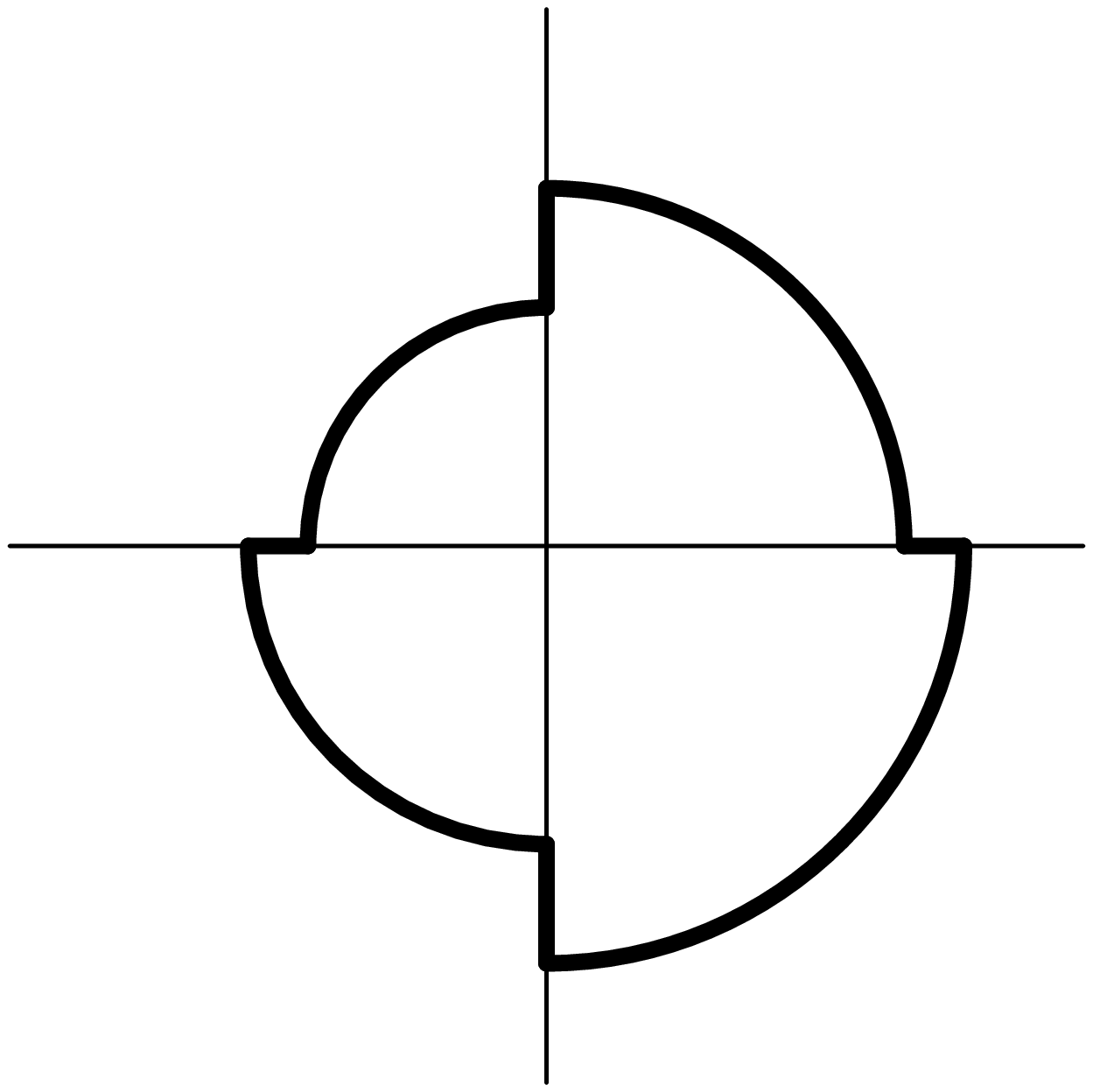}}}
\caption{\small Left: A road network spanning an $6\times6$ km$^2$
region. Right: An instance of AP's coverage
region with its boundary highlighted.} \label{fig:setting}
\end{figure}

Figure~\ref{fig:setting}(left) shows the road network used in our study. The network has 1802 road intersections and 2377 road segments. We assume each road segment has two lanes in the opposite directions and ignore the width of lanes.  The travel speed of each segment is in the interval [10m/s,20m/s]. 
Each road intersection is a candidate location for deploying APs with a data rate in the interval [5Mbps,10Mbps]. The coverage region at each candidate
location is modeled using a sector based approach from~\cite{sector}, where each region is
composed of 4 sectors of $90^{\circ}$ with radius randomly selected from [150m,250m], as shown in Figure~\ref{fig:setting}(right).
Except in the two-stage setting discussed in Section~\ref{sec:simulation:twostage}, each AP has a unit cost.
The set of movements $P$ consists of 10000 paths randomly sampled from all the shortest paths of
length at least 2km connecting two road intersections. 
For Algorithm~\ref{alg:search}, the parameter $\delta$ used in
the binary search is set to $0.0005$, and for Algorithm~\ref{alg:bic2}, the parameter $\tau$ is set to $0.01$.

To simulate the traffic density on each road segment, we generate a movement file with 1000 mobile users moving in the network for 24 hours. A restricted random waypoint mobility model is considered. A user starts at a randomly selected road intersection $a$, and randomly picks another road intersection $b$ of distance at least 2km away from $a$, and moves to $b$ by following the shortest path connecting the two intersections. After reaching $b$, the user immediately picks a new destination $c$ of 2km away, and moves towards $c$, and so on. The travel speed on each road segment is sampled from the corresponding interval. We then estimate the user density on each segment from the movement file.

\subsubsection{Contact Opportunity in Distance}
We first study Algorithm~\ref{alg:bic} for minimum cost contact opportunity in distance (MinCost for short), and Algorithm~\ref{alg:search} for maximum contact opportunity in distance with a budget (MaxOpp for short). We compare our algorithms with the following two baseline algorithms, where $\hat{A} \subseteq A$ denotes the set of coverage regions that touch at least one path in $P$:

\begin{enumerate}
\item {\bf Uniform random sampling} (Rand for short), which at
each step randomly picks a new element from $\hat{A}$ until the required contact opportunity is obtained (for the minimum cost problem {\bf P1}), or until the budget is reached (for the maximum coverage problem {\bf P2}).
\item {\bf Max-min distance sampling}~\cite{distance-sampling}
(Dist for short), which starts at a randomly selected location in
$\hat{A}$, and at each step finds a new element from $\hat{A}$ that
maximizes the minimum graph distance (in terms of shortest paths)
from the elements already selected, until the required contact opportunity is obtained (for {\bf P1}), or until the budget is reached (for {\bf P2}).
\end{enumerate}

Note that both algorithms involve randomness. In the simulation, each of them is repeated 100 times.

\begin{figure}
\centering
\begin{tabular}{cc}
\includegraphics[width=1.65in]{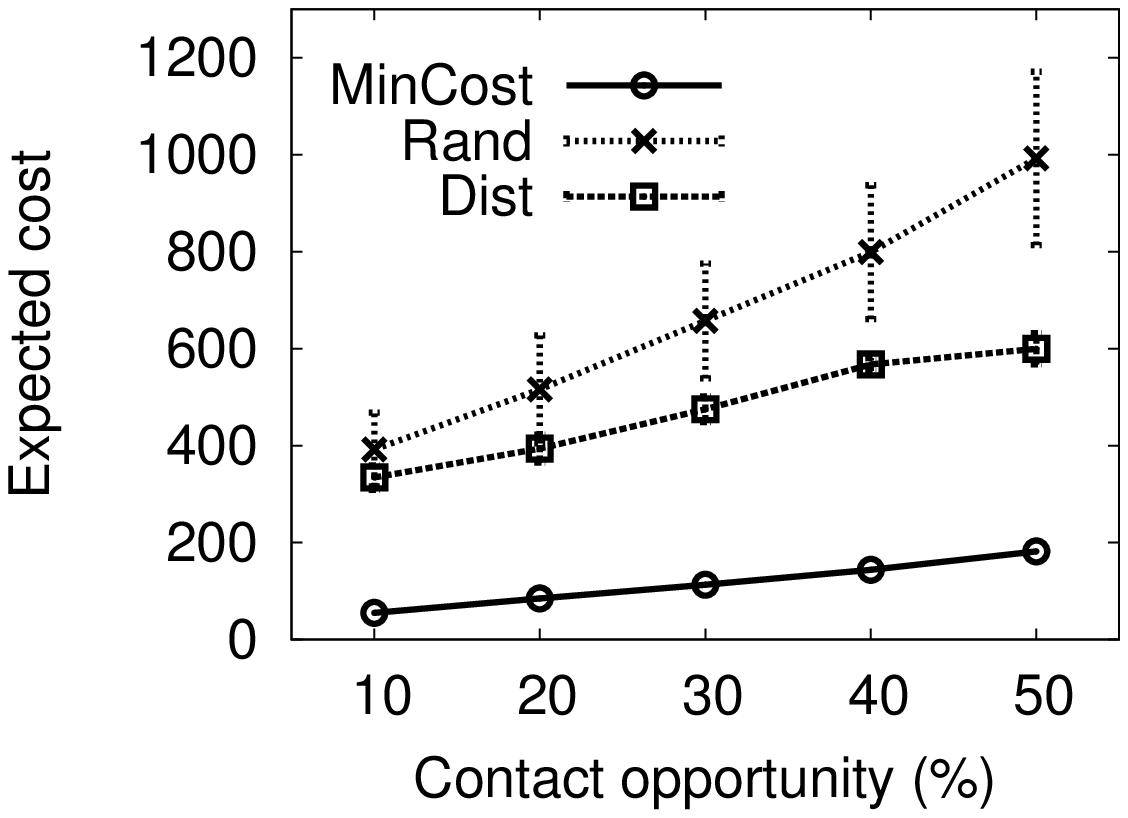} &
\includegraphics[width=1.65in]{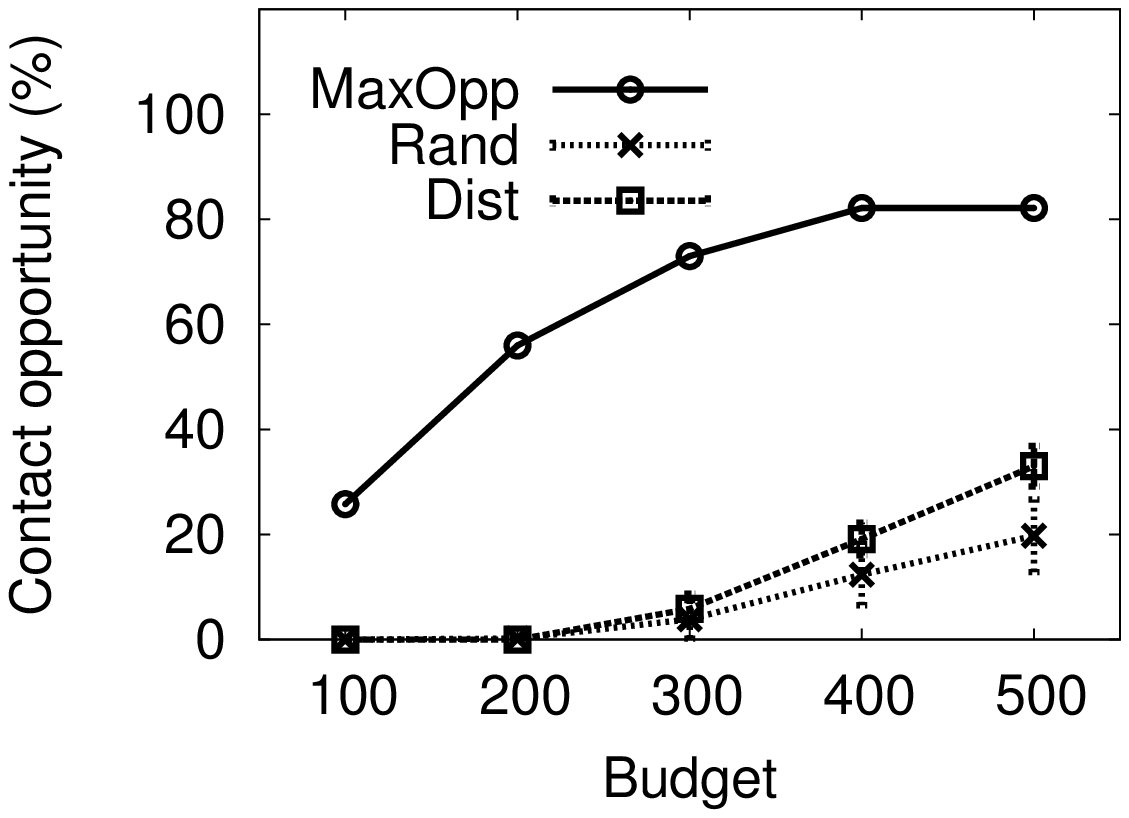} \\ 
 {\small (a)} &
 {\small (b)} \\ 
\end{tabular}
\caption{\small (a) Cost for achieving a required level of contact opportunity in distance. (b) Minimum contact opportunity across all the movements for a given budget. 
} \label{fig:dist}
\end{figure}

Figure~\ref{fig:dist}(a) shows the expected cost for ensuring a required level of contact opportunity in distance for the three algorithms. The error bars represent the standard deviations across all the deployments generated by the baseline algorithms. We observe that our algorithm reduces the cost to 15\% - 30\% of the baseline cost while achieving the same level of contact opportunity. Also note that the random sampling technique performs worst among the three algorithms and has a large standard deviation.

Figure~\ref{fig:dist}(b) shows the minimum contact opportunity in distance (across the set of paths) that can be achieved under various budgets, where the budget is simply the number of APs allowed to use since each candidate location has a unit cost. Our algorithm achieves more than 200\% higher contact opportunity in all the cases. In fact, when the budget is low, the minimum contact opportunity of the two baseline algorithms is very close to 0. 
Moreover, using about 400 APs, our algorithm achieves a value very close to what can be achieved by a full deployment, that is, deploying one AP at each of the 1802 candidate locations (not shown in the figure). 

\begin{figure}[!ht]
\centering
\begin{tabular}{cc}
\includegraphics[width=1.65in]{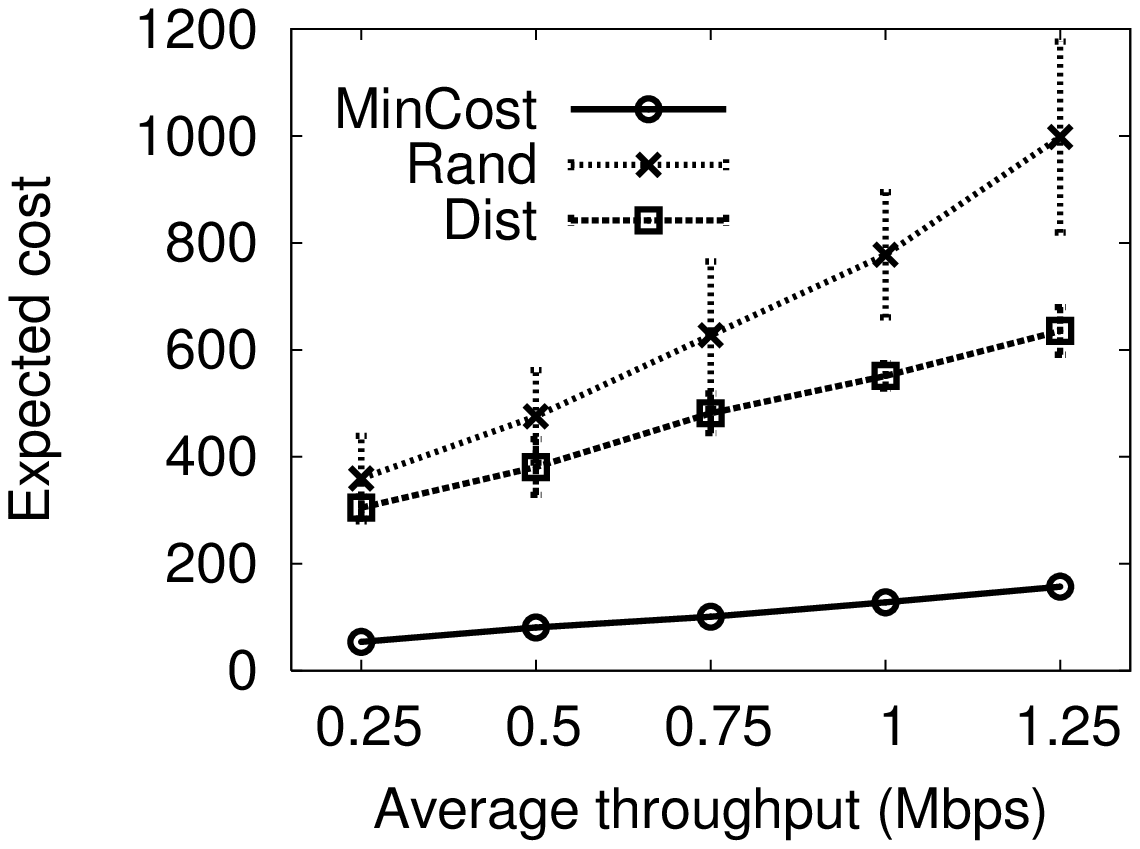} &
\includegraphics[width=1.65in]{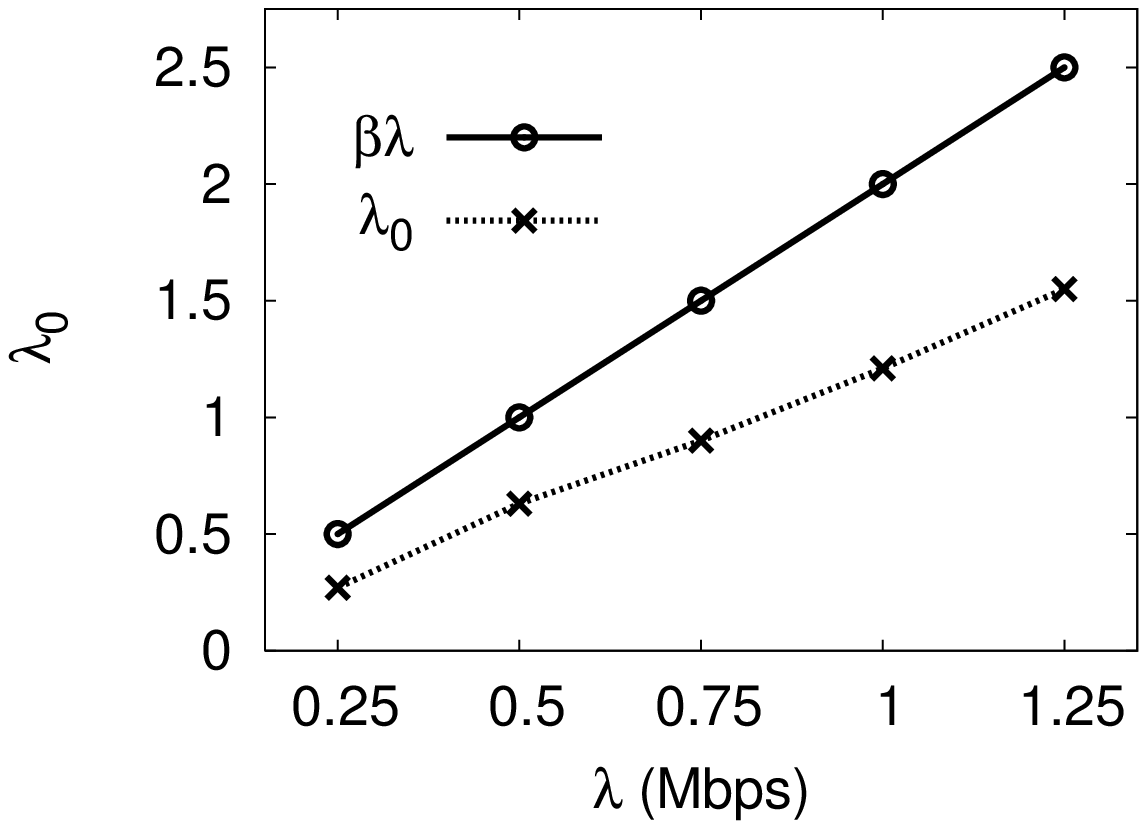} \\ 
 {\small (a)} &
 {\small (b)} \\ 
\end{tabular}
\caption{\small (a) Cost for achieving a required average throughput (across all the movements and all the scenarios). (b) Minimum $\lambda_0$ for getting a feasible solution in Algorithm~\ref{alg:bic2}. 
} \label{fig:robust}
\end{figure}

\begin{figure*}[!ht]
\centering
\begin{tabular}{cccc}
\includegraphics[width=1.65in]{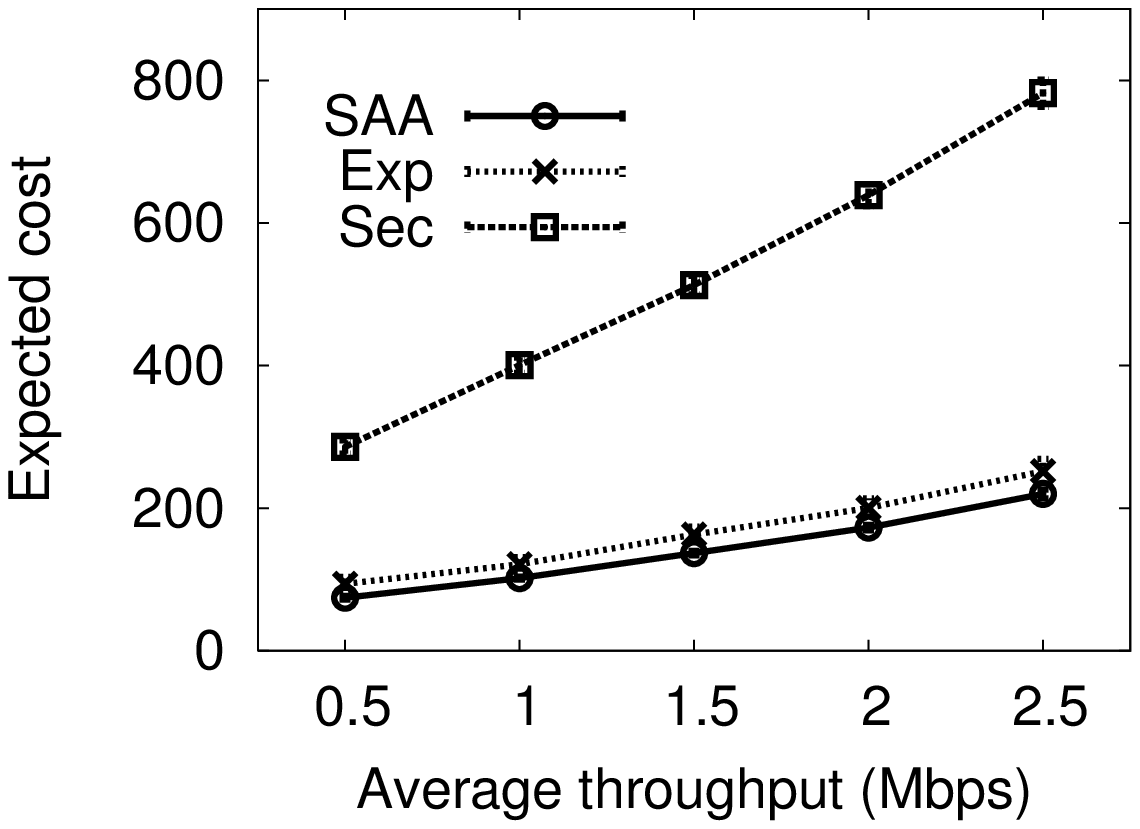} &
\includegraphics[width=1.65in]{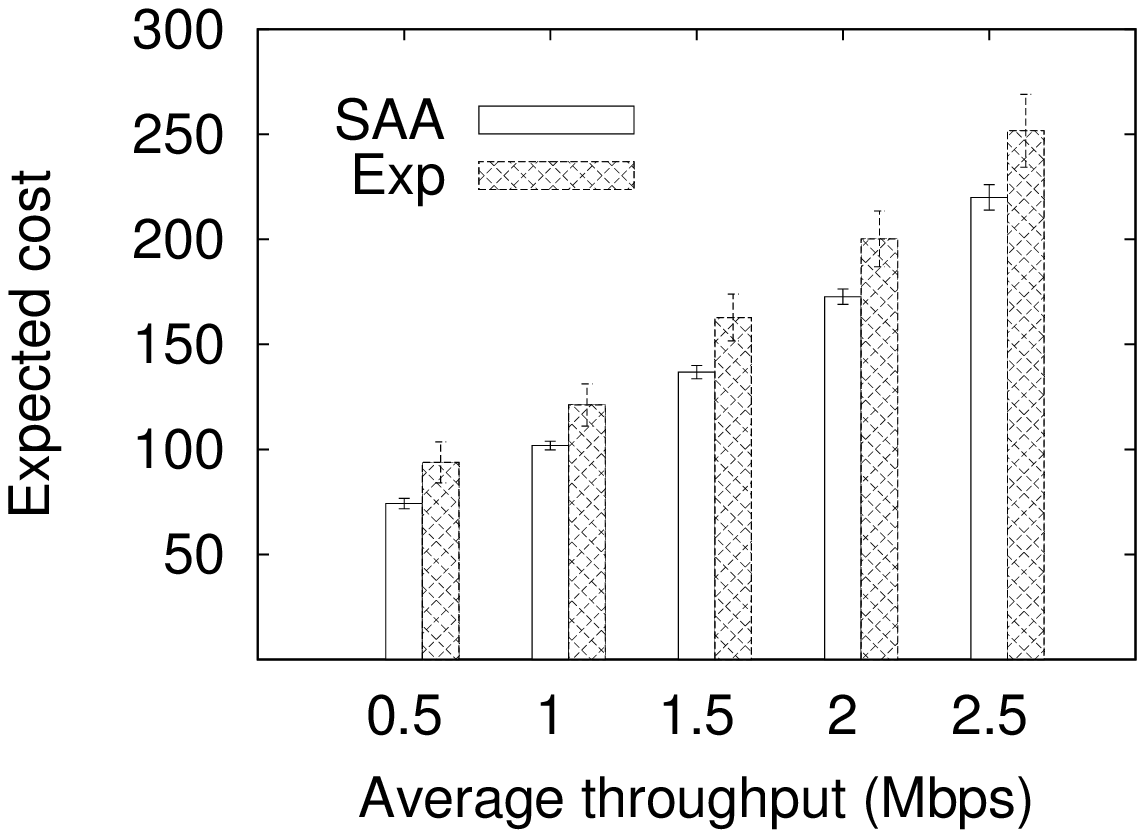} &
\includegraphics[width=1.65in]{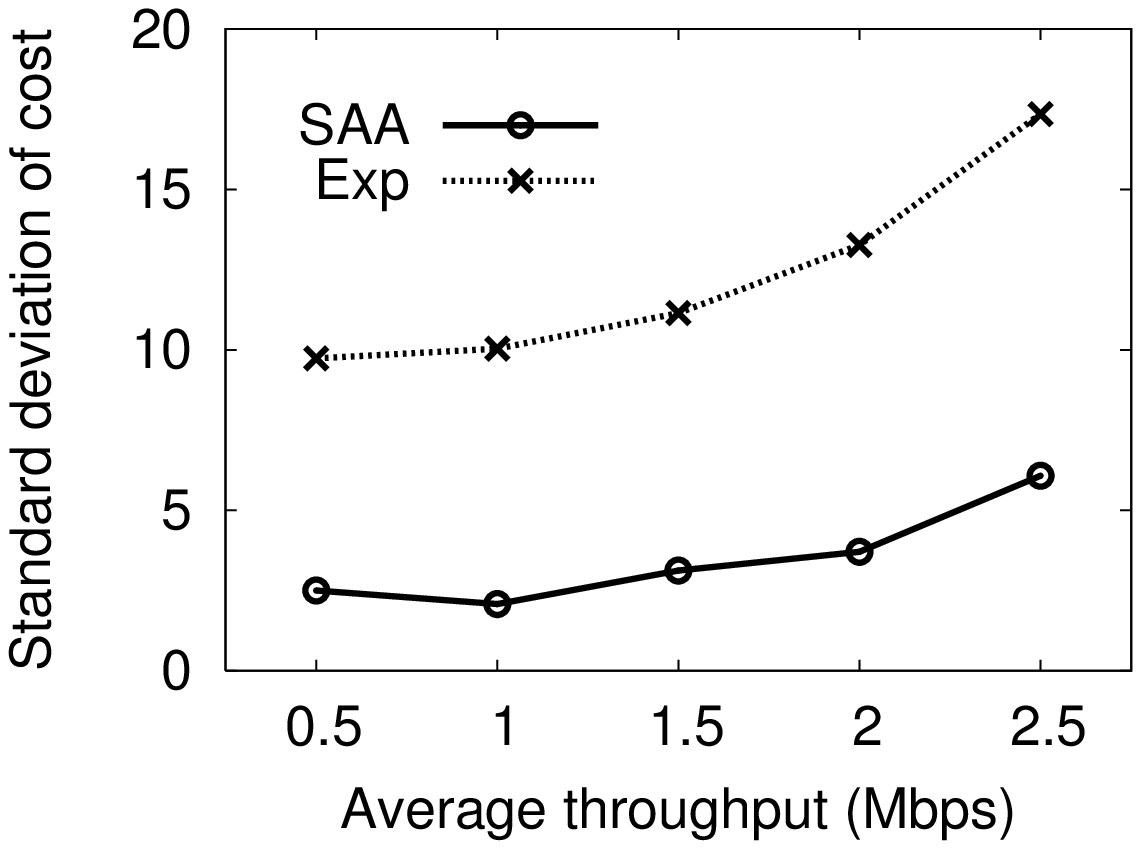} &
\includegraphics[width=1.65in]{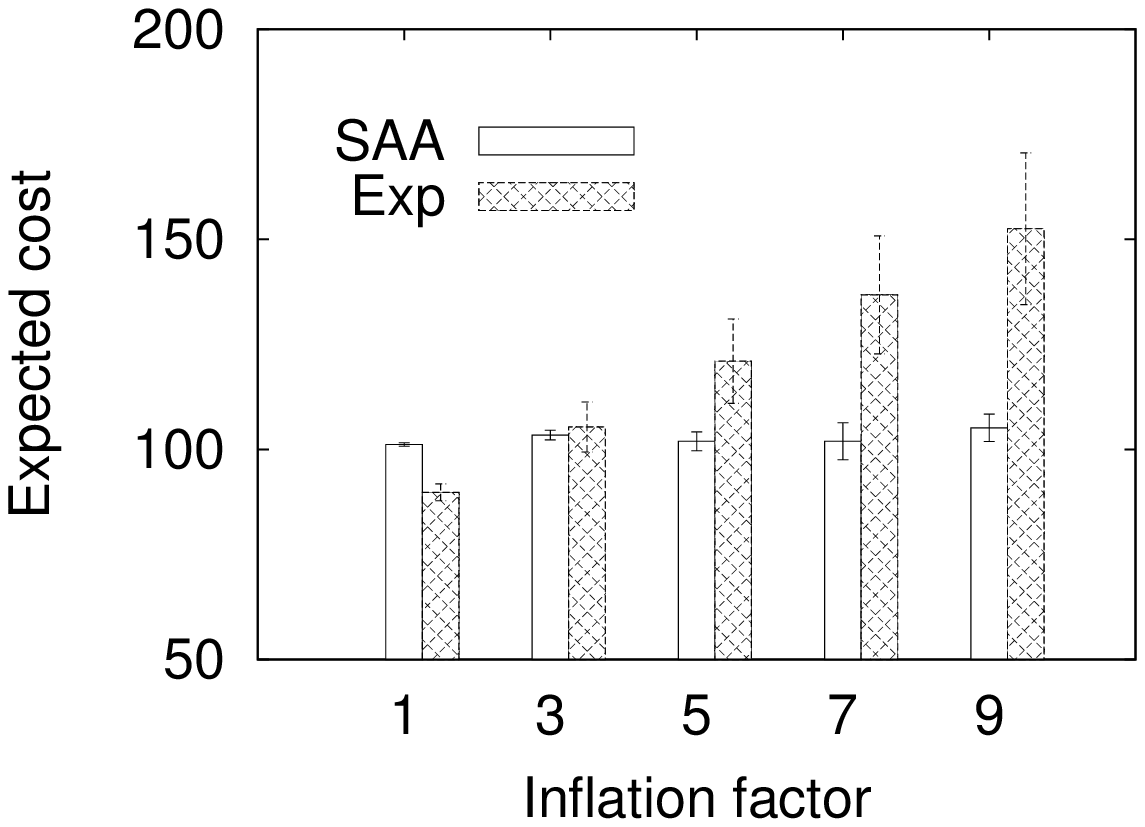} \\
 {\small (a) } &
 {\small (b) } &
 {\small (c) } &
 {\small (d) }
\end{tabular}
\caption{\small (a) Total cost for a required average throughput in two-stage deployment. (b) Total cost for SAA vs. Exp. (c) Standard deviation of total cost for SAA vs. Exp. (d) Total cost for a required average throughput of 1Mbps under various inflation factors.} \label{fig:two}
\end{figure*}

\subsubsection{Robust Average Throughput Optimization}
We then study the performance of Algorithm~\ref{alg:bic2} for {\bf P3} for achieving robust average throughput under uncertainty. The performance of the ``mean speed'' scenario based algorithm for {\bf P4} will be studied in ns-3 based simulations. The two baseline algorithms used before are extended as follows. In each iteration, the baselines apply Algorithm~\ref{alg:scenario} to check if the partial solution has already guaranteed the required average throughput in the worst-case scenario. 

Figure~\ref{fig:robust}(a) shows the expected cost for achieving a required average throughput for a mobile user moving through any path in $P$ and under any scenario, where the error bars again denote the standard deviations. We observe that our algorithm reduces the cost to less than 25\% of the baseline cost, and random sampling again performs worst among the three algorithms. Figure~\ref{fig:robust}(b) shows the minimum value of $\lambda_0$ in Algorithm~\ref{alg:bic2} when the solution first becomes feasible (line 7 in the algorithm). As we shown in Section~\ref{sec:robust}, such a $\lambda_0$ is upper bounded by $\beta \lambda$. Figure~\ref{fig:robust}(b) verifies this result with $\beta = 20/10 = 2$. Moreover, it shows that $\lambda_0$ is actually bounded by $1.25\lambda$ in the simulation setting; hence, Algorithm~\ref{alg:bic2} has a better performance than the theoretical bound. 

\subsubsection{Two-stage Stochastic Optimization}\label{sec:simulation:twostage}
Finally, we study the performance of the SAA based algorithm (SAA for short) for minimizing the total cost for achieving a required average throughput in the two-stage setting. The scenario distribution is generated by assuming $v_e,h_e$, and $r_a$ are independently and uniformly distributed in the corresponding intervals for all the road segments and all the APs. Each AP has a unit first stage cost, and a second cost  determined by an inflation factor. We first generate 2000 samples of scenarios, and use 1000 of them as learning samples for the SAA based method, that is, the initial deployment is found for these samples using the polynomial-scenario extension of Algorithm~\ref{alg:bic} presented in Section~\ref{sec:two-stage}. Note that the sample size is relatively small compared with the network size and the number of movements considered. The rest 1000 samples are then used for testing, where in each scenario, the initial deployment is supplemented to meet the throughput requirement. This algorithm is compared with the following two heuristics:

\begin{enumerate}
\item {\bf Expected scenario} (Exp for short), which is discussed in Section~\ref{sec:two-stage}, where the initial deployment is found by directly applying Algorithm~\ref{alg:bic} to the ``mean'' scenario, which is then augmented for each of the 1000 testing samples using Algorithm~\ref{alg:bic}.
\item {\bf Second stage only} (Sec for short), which does not consider the first stage, and a new deployment is found for each testing sample using Algorithm~\ref{alg:bic}.
\end{enumerate}

We first consider a fixed inflation factor of $5$ (hence each AP has a fixed second stage cost of 5). Figure~\ref{fig:two}(a) shows the total cost for achieving a required average throughput. The second stage only approach is clearly the worst among the three algorithms due to the high cost of the second stage. To see the performance of SAA and Exp clearly, their total cost and standard deviations are replotted in Figures~\ref{fig:two}(b) and (c), respectively. We observe that Exp performs 15\%-25\% worse than SAA and suffers from a high standard deviation. Figure~\ref{fig:two}(d) further illustrates the performance of SAA and Exp for different inflation factors. We observe that SAA performs worse only when the inflation factor is close to 1. Actually, when the inflation factor is 1, that is, the two stages have the same cost, there is no benefit to have an initial deployment, and hence the second stage only approach is the best (not shown in the figure). For large inflation factors, SAA always performs better than Exp and the reduction in cost increases as the inflation factor becomes larger, which highlights the deficiency of using the ``mean" scenario cost to estimate the expected cost. Moreover, SAA has a stable performance under different inflation factors and a small standard deviation.

\begin{figure*}[!ht]
\centering
\begin{tabular}{cccc}
\includegraphics[width=1.65in]{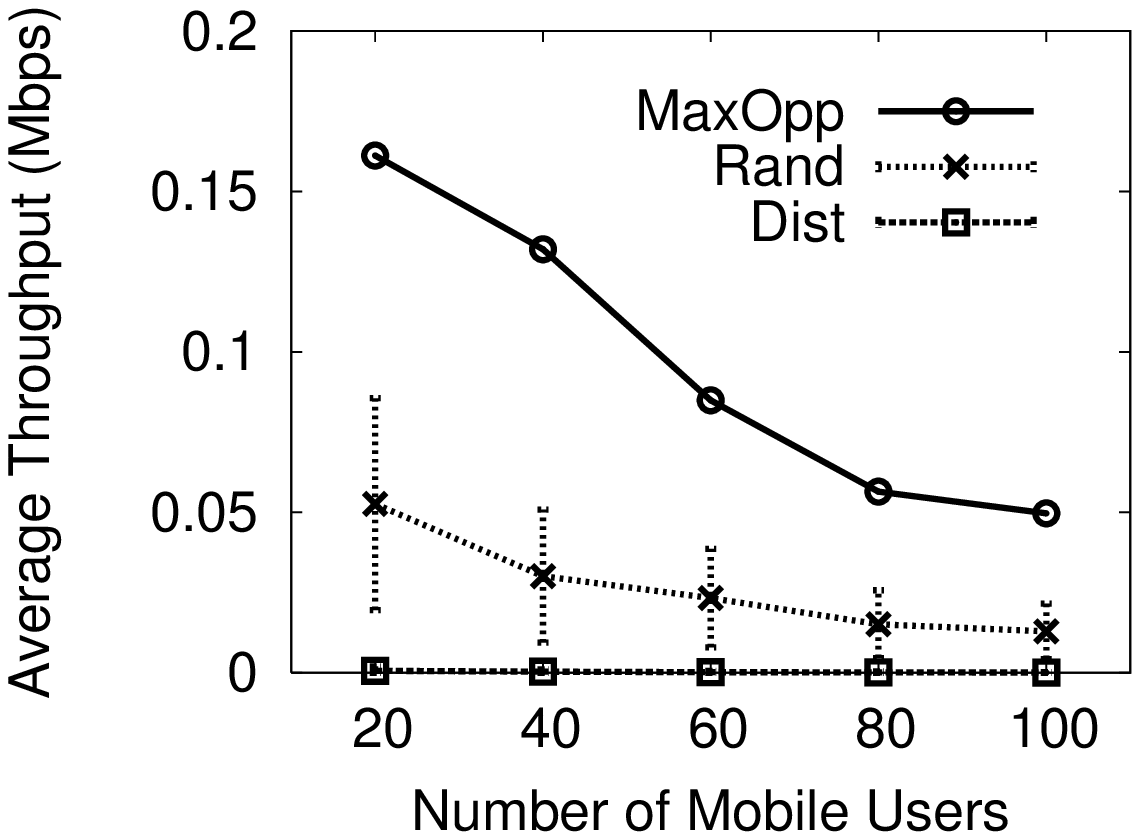} &
\includegraphics[width=1.65in]{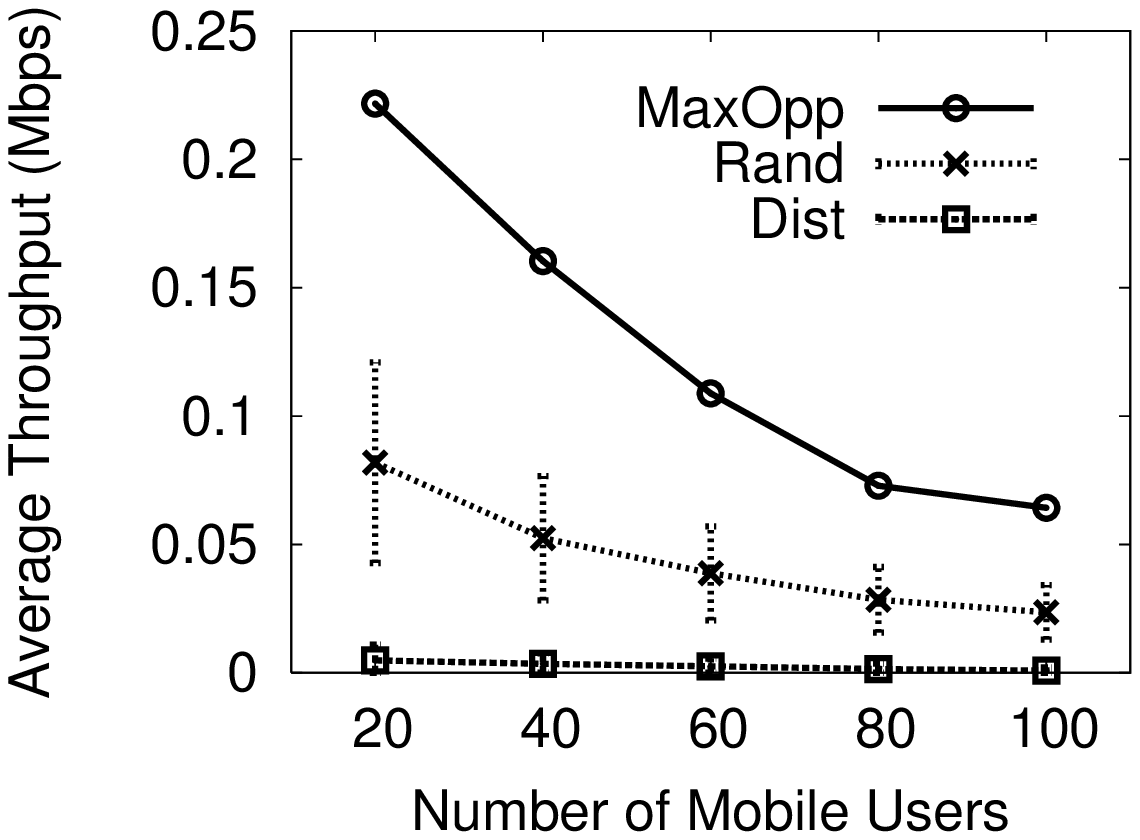} &
\includegraphics[width=1.65in]{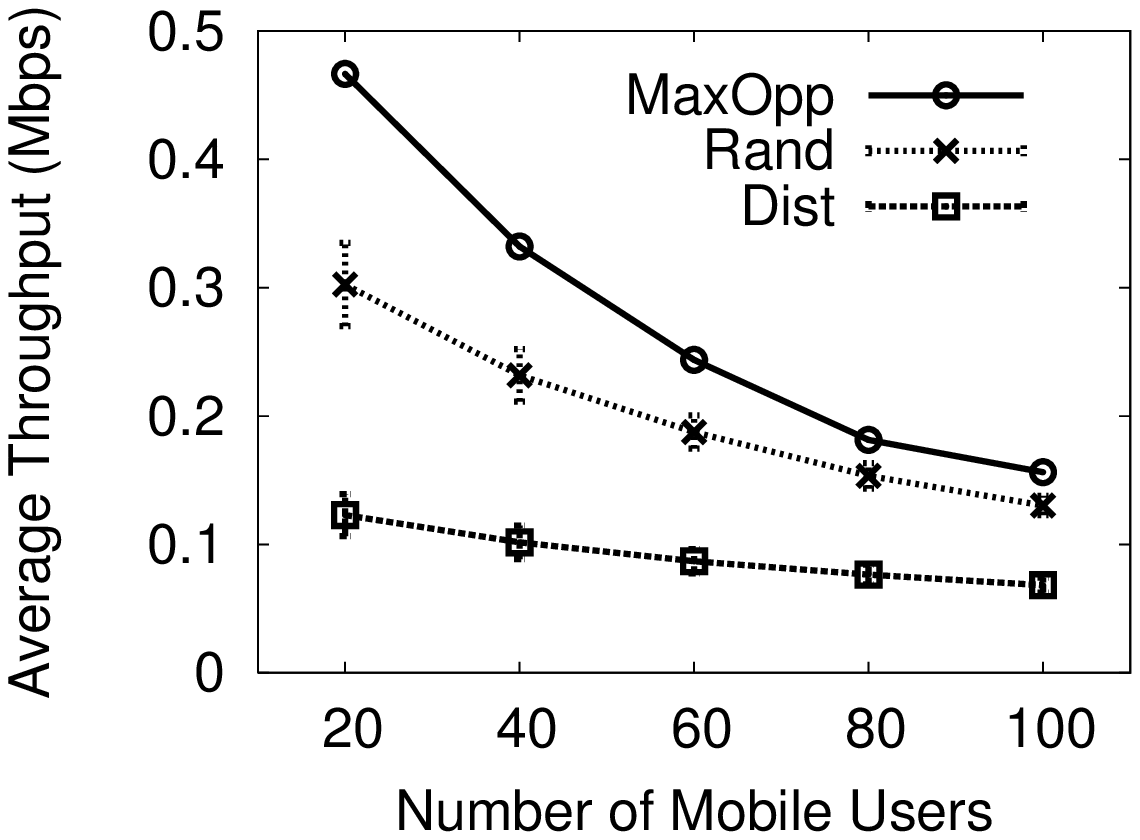} &
\includegraphics[width=1.65in]{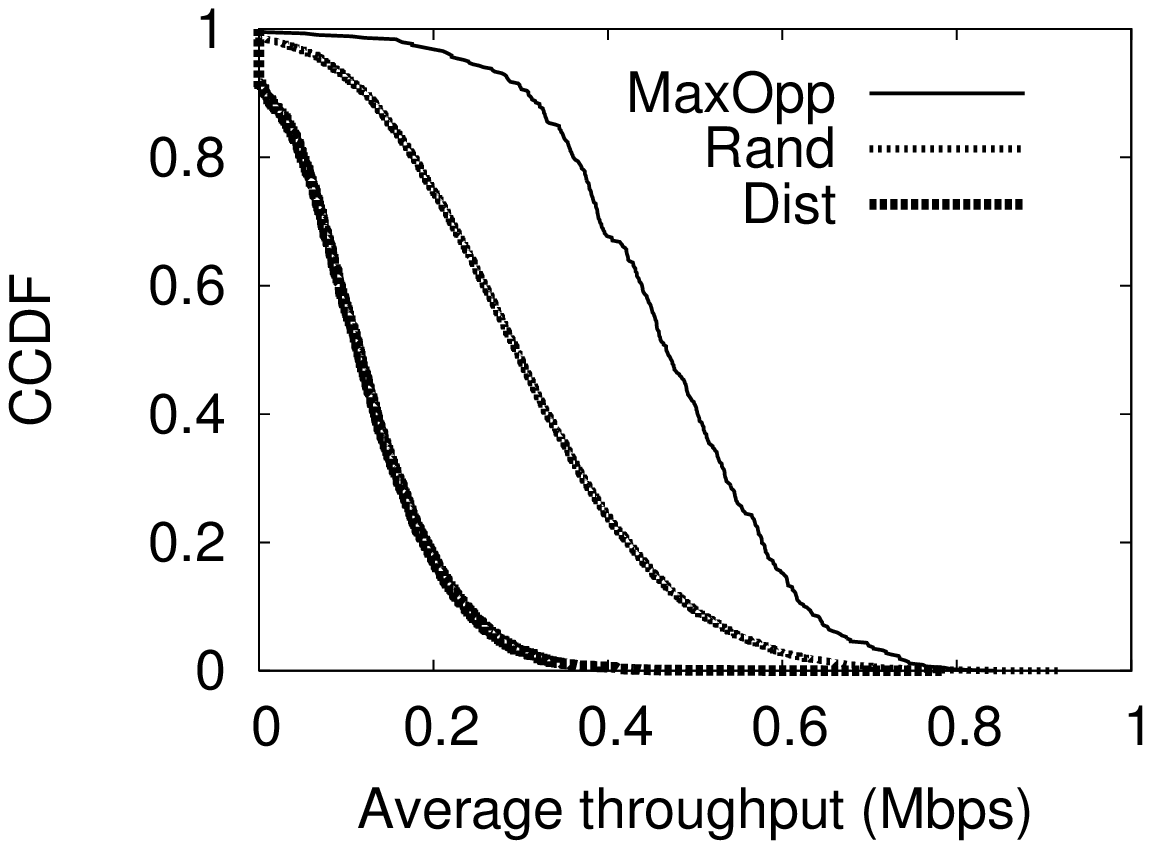} \\ 
 {\small (a) } &
 {\small (b) } &
 {\small (c) } &
 {\small (d) } \\
\end{tabular}
\caption{\small (a) Average throughput of the worst 5\% paths vs. budget. (b) Average throughput of the worst 10\% paths vs. budget. (c) Average throughput of all the paths vs. budget. (d) CCDF of average throughput across all the paths and deployments (20 mobile users).} \label{fig:ns3}
\end{figure*}

\subsection{Ns-3 Simulations}
We then conduct ns-3 based packet level simulations to further study the performance of our algorithms. Note that contact opportunity in distance is independent of packet transmission, and the objective of cost minimization is hard to simulate. Therefore, our focus is on maximization throughput for a given budget under a randomly generated traffic scenario.

\subsubsection{Simulation Setting}
Due to the high overhead for simulating large scale mobility and data transmission in ns-3, we use a smaller road network (a 2km $\times$ 2km subregion in the same area as the large network with the same travel speed distribution). We fix the number of APs at 20 and vary the number of mobile users, denoted as $K$, between 20 and 100. For each $K$, we first generate a 24 hour movement file with $K$ users as before. The movement file is then used to estimate the user density. We then run our ``mean speed'' scenario based algorithm for problem P4 to generate a deployment for each $K$, and the random sampling algorithm and the distance sampling algorithm to generate 20 deployments each.

In each simulation, 20 static nodes are set up as APs with their locations determined by a deployment file, and $K$ mobile nodes are generated with their mobility determined by the movement file. The set of nodes are configured as follows. In the physical layer, we use the constant speed propagation delay model with the default speed (the speed of light), and the Friis propagation loss model~\cite{Friis-model}. We have extended the loss model to allow four different energy thresholds that match the communication ranges in the four directions as in Figure~\ref{fig:setting} (right). All the ranges are randomly sampled from the interval of [150m, 250m] as before. In the MAC layer, 802.11g protocol is used with a constant data rate of 6 Mbps. Each AP has a different SSID, and APs that are close to each other are assigned different channels to avoid interference (ns-3 WiFi does not model cross-channel interference). Each mobile node is configured with multiple channels so that it can download data from any APs in range, but the association protocol ensures that a node is associated with at most one AP at any time. In the application layer, CBR traffics are generated from each AP to mobile users served by it. To reduce communication overhead, mobile nodes do not actively probe channels. They only wait for beacons from APs. 
Whenever a node encounters a new AP or is disassociated from an old AP, it chooses from the set of APs in range the one with the least number of users associated, where the tie is broken by giving higher priority to the newly encountered AP. An AP serves all the nodes associated with it in an equal data rate with the total rate bounded by 1 Mbps.

\subsubsection{Simulation Result}
In Figure~\ref{fig:ns3}(a), the average throughput for the 5\% of paths of minimum throughput is plotted, where the average is taken over all these paths and over all the deployments. Figure~\ref{fig:ns3}(b) shows the similar results for 10\% of worst paths. In both cases, our algorithms achieves more than 150\% of higher performance. 
Moreover, Figure~\ref{fig:ns3}(c) shows that although our algorithm is designed to optimize the worst-case performance, it also achieves significant higher throughput in the average sense, where the average throughput over all the paths is plotted. Figure~\ref{fig:ns3}(d) plots the complementary cumulative distribution of throughput across all the paths for the 20 mobile user case. The figure illustrates that our algorithm not only achieves a better worst-case and average performance, but also dominates the baselines in the stochastic sense (roughly). In addition, we find that distance sampling performs even worse than random sampling in this setting, which is contrast to the case of cost minimization as we observed before. One explanation is that distance sampling distributes APs in a more uniform way and when the budget is low, it does not provide enough coverage to short movements.

\section{Experimental Evaluation}\label{sec:exp}
We set up a small scale controlled experiment to better understand
the performance of our approach. The experiment was carried out in a
180m $\times$ 120m parking lot located at the west campus of OSU and
is free of potential interference from other WiFi networks. The
experiment was usually carried out at night when the parking lot was
empty. We artificially divided the parking lot area into a 6 by 4
grid and use it as a small road network. All the 24 intersections
are treated as candidate locations for deploying APs.

A single mobile node carried by a car and 4 APs are used in the
experiment. Each AP is a laptop equipped with an Orinoco 802.11b/g
PC card and an external antenna mounted on a 1.7m high tripod so
that the signal will not be blocked by the car in the test. The
single mobile node is a laptop equipped with a Ubiquiti Networks SRC
802.11a/b/g PC card and two external antennas fixed at the two sides
of the car. The transmission power of each AP is set to 6 dBm, which
is tested to give an effective transmission distance of no more than
50 meters. Each node runs Ubuntu Linux with Linux 2.6.24 kernel and
madwifi device driver for the 802.11 interface. The physical layer
data rate of each node is fixed at 54Mbps.

\begin{figure}[!h]
\centering
\includegraphics[width=2.8in]{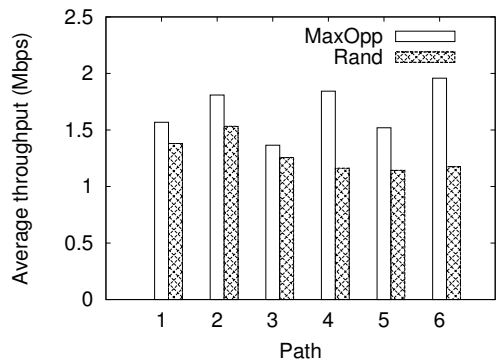}
\caption{\small The average throughput of the 6 paths under
evaluation, where Rand represents the average of 5 random
deployments. } \label{fig:path}
\end{figure}

A total of 5 random deployments are evaluated and compared with a
deployment computed by Algorithm~\ref{alg:search} for maximizing the contact opportunity in distance across the set of shortest paths
between intersections of length at least 200m (there are 30 such
paths in total), with a budget 4. The algorithm assumes that each AP has a unit cost
and the coverage region of each AP is a disk of a radius 50m.

Because of the large volume of driving work and limited availability
of that place, we picked 6 representative shortest paths that go
through different parts and directions of the parking lot, and drove
through each of them 3 times for each deployment. The moving speed
is kept at about 10mph. When moving through a path, the mobile node
attempts to associate with an AP with the strongest signal. Once
associated, it downloads UDP packets from the AP until it is
disconnected from that AP. The mobile node then finds another AP
with the strongest signal to associate. Figure~\ref{fig:path} shows
the average throughput of each of the 6 paths. For random sampling,
the average results across the five deployments are plotted. We
observe that our solution achieves up to $66.7\%$ higher throughput,
and across all the 6 paths the average improvement in throughput is
$26.4\%$. These results are promising and serve as a first step towards larger scale prototype deployment. Compared with simulations, the improvement is less significant due to: (1) the instability of channel condition in the outdoor environment; (2) the overhead of association and disassociation; and (3) the small driving area, where even 4 APs have a high chance of covering almost the entire area in a random deployment. We will investigate these issues in our future work to further improve the performance.

\ignore{
\begin{figure}[!h]
\centering
\includegraphics[width=3in]{figures/exp.eps}
\caption{\small The average throughput of the 6 paths under
evaluation, where Rand represents the average of 5 random scenarios.
} \label{fig:path}
\end{figure}}

\section{Conclusion}\label{sec:concl}
Existing solutions for wide-area data services fail to provide an
economically scalable infrastructure for serving mobile users with QoS guarantees. This paper proposes a systematic approach to address this fundamental problem.
We first present Contact Opportunity
as a new metric for measuring roadside WiFi networks. We then discuss efficient deployment techniques for minimizing the cost for ensuring a required level of contact
opportunity, and for maximizing the achieved contact opportunity under a budget. We further extend this metric to
average throughput under uncertainty, which is more intuitive for mobile users and
application designers, and study deployment techniques for minimizing the worst-case cost and the expected cost, respectively, for ensuring a required average throughput.
Using simulations and experiments, we show that our approach achieves a much better cost vs. throughput tradeoff compared with some
commonly used deployment techniques.

\section{Acknowledgment}
This material is based upon work supported by NSF grants CNS-0403342, CNS-0721817, CNS-0721983, CCF-0728928, CNS-0910878, CNS-1254525,
and Fedex Institute of Technology (FIT) at the University of Memphis.

\appendix

\noindent {\bf Proof of Proposition~\ref{prop:scenario}}: To simplify the notation, we use $i, j, k$, etc., to denote road segments (edges). For an edge $i$, define $t^1_i = d_i/v^1_i$ and $t^2_i = d_i/v^2_i$. Then the travel time over $i$, denoted as $t_i$, is within $[t^2_i, t^1_i]$. To prove the first statement, we take the partial derivative of $\gamma_p$ over $t_i$ in \eqref{equ:throughput2}:
\begin{align}
\frac{\partial \gamma_p}{\partial t_i} & = \frac{r_i\sum_k t_k - \sum_k r_k t_k}{(\sum_k t_k)^2} \nonumber \\
& = \frac{\sum_{k \neq i} t_k(r_i - r_k)}{(\sum_k t_k)^2} \label{equ:derivative}
\end{align}

\noindent Since the partial derivative does not depend on $t_i$, for fixed $r_i, r_k, t_k$, $\gamma_p$ is minimized by setting $t_i = t^1_i$ if $\frac{\partial \gamma_p}{\partial t_i} \leq 0$, and $t_i = t^2_i$ if $\frac{\partial \gamma_p}{\partial t_i} > 0$. To prove the second statement, consider an optimal assignment $t^*_k, \forall k \in E_p$ that minimizes $\gamma_p$. We need to show that for any pair $i, j \in E_p$, if $r_i > r_j$, and $t^*_j = t^2_j$, then $t^*_i = t^2_i$. Since $t^*_j = t^2_j$, we must have $\frac{\partial \gamma_p}{\partial t_j} \geq 0$ under the optimal assignment, or equivalently, $\sum_{k \neq j} t^*_k(r_j - r_k) \geq 0$ by \eqref{equ:derivative}. We will show that $\frac{\partial \gamma_p}{\partial t_i} > 0$ under the optimal assignment, and hence $t^*_i = t^2_i$. Note that $\sum_{k \neq j} t^*_k(r_j - r_k) = A + t^*_i(r_j-r_i)$ where $A = \sum_{k \not \in \{i,j\}} t^*_k(r_j - r_k)$, and $\sum_{k \neq i} t^*_k(r_i - r_k) = A' + t^*_j(r_i-r_j)$ where $A' = \sum_{k \not \in \{i,j\}} t^*_k(r_i - r_k)$. Since $r_i > r_j$, we have $A' > A, t^*_i(r_j-r_i) < 0, t^*_j(r_i-r_j)>0$. It follows that $\sum_{k \neq i} t^*_k(r_i - r_k) > \sum_{k \neq j} t^*_k(r_j - r_k) \geq 0$. Hence $\frac{\partial \gamma_p}{\partial t_i} > 0$ as we require. \hspace*{\fill}~\IEEEQED\par

\vspace{2ex}
\noindent {\bf Proof of Proposition~\ref{prop:nonsubmodular}}:
It is clear that $\gamma_p(S,k_S) = 0$ when $S = \emptyset$. To prove the monotonicity, consider any two subsets $S \subseteq T \subseteq A$.
For any scenario $k$, we have $r_e(S) = (\sum_{l \in L_e \cap L_S} d_l r_l)/d_e \leq (\sum_{l \in L_e \cap L_T} d_l r_l)/d_e = r_e(T)$. Thus we have $\gamma_p(S,k) \leq \gamma_p(T,k)$ by definition \eqref{equ:throughput2} for any scenario $k$. Therefore, $\gamma_p(S,k_S) \leq \gamma_p(S,k_T) \leq \gamma_p(T,k_T)$ where the first inequality follows from the fact that $k_S$ is a worst-case scenario for $S$.

To see that $\eta^t_p$ is not submodular, it suffices to give a
counter-example. Consider a toy road network with a single road segment corresponding to the line segment from $(0,0)$ to $(3,0)$ in
$\mathbb{R}^2$ with two road intersection $(0,0)$ and $(3,0)$, and the single path connecting them. There are three candidate
locations at points $v_i = (i+0.5,0), i=0,1,2$, where each of them covers
an interval $[(i,0),(i+1,0)]$, and $p$ is
partitioned into 3 subsegments by them. Each subsegment has a travel speed within $[0.5,1]$ and a single mobile user. Each AP has a unit data rate. Let $S = \{a_1\}$, $T =
\{a_1,a_2\}$. Then $S \subseteq T, a_3 \not
\in T$. $\gamma_p(S) = 1/(2 \times 2+1) = 0.2, \gamma_p(S \cup
\{a_3\}) = \gamma_p(T) = 2/(1 \times 2+2) = 0.5$, and $\gamma_p(T \cup
\{a_3\}) = 1$. Therefore, $\gamma_p(S \cup \{a_3\}) - \gamma_p(S) =
0.3$, while $\gamma_p(T \cup \{a_3\}) - \gamma_p(T) = 0.5$. Hence the submodularity does not hold.\hspace*{\fill}~\IEEEQED\par

\vspace{2ex}
\noindent{\bf Proof of Proposition~\ref{theorem:robust}}: The first inequality is clear by the definition of $k_S$. To show the second inequality, let $t^1_e = d_e/v^1_e$ and $t^2_e = d_e/v^2_e$. By the assumption, $t^1_e/t^2_e \leq \beta$ for all $e$. Define $R = \sum_{e \in E_p} r_e(t^1_e+t^2_e)/2, T = \sum_{e \in E_p} (t^1_e+t^2_e)/2$. Then $\gamma_p(S,k_0) = \frac{R}{T}$.
Let $E_1 \subseteq E_p$ denote the set of edges that take the low speed in $k_S$, and $E_2 \subseteq E_p$ the set of edges that take the high speed in $k_S$. Define $R' = \sum_{e \in E_1} r_et^1_e+\sum_{e \in E_2} r_et^2_e$, and $T' = \sum_{e \in E_1} t^1_e+\sum_{e \in E_2} t^2_e$. Then $\gamma_p(S,k_S) = \frac{R'}{T'}$. We further have
\begin{align*}
\frac{R}{R'} & = \frac{\sum_{e \in E_p} r_e(t^1_e+t^2_e)/2}{\sum_{e \in E_1} r_et^1_e+\sum_{e \in E_2} r_et^2_e} \\
& = \frac{1}{2}\bigg(1+\frac{\sum_{e \in E_1} r_et^2_e+\sum_{e \in E_2} r_et^1_e}{\sum_{e \in E_1} r_et^1_e+\sum_{e \in E_2} r_et^2_e}\bigg)\\
& \leq \frac{1}{2}\bigg(1+\frac{\sum_{e \in E_p} r_et^1_e}{\sum_{e \in E_p} r_et^2_e}\bigg) \\
& \leq \frac{1}{2}(1+\beta)
\end{align*}

\noindent Similarly, we can show that $\frac{T'}{T} \leq \frac{2\beta}{1+\beta}$. Hence $\frac{\gamma_p(S,k_0)}{\gamma_p(S,k_S)} = \frac{R}{R'}\frac{T'}{T} \leq \beta$. \hspace*{\fill}~\IEEEQED\par 

\bibliographystyle{abbrv}
\bibliography{reference}
\end{document}